\documentclass[aps,pre,twocolumn,a4paper,showpacs]{revtex4}

\usepackage{amsmath}
\usepackage{graphicx}
\usepackage[pdftex]{hyperref}
\hypersetup{colorlinks=true,linkcolor=blue,citecolor=blue,urlcolor=blue}

\renewcommand{\vec}[1]{\mathbf{#1}}
\renewcommand{\Re}{\operatorname{Re}}
\newcommand{\sign}{\operatorname{sign}}

\begin{document}

\title{Glass-like behavior of a hard-disk fluid confined to a narrow channel}
\author{J. F. Robinson}
\author{M. J. Godfrey}
\author{M. A. Moore}

\affiliation{School of Physics and Astronomy, University of
  Manchester, Manchester M13 9PL, United Kingdom}

\date{\today}

\begin{abstract}
Disks moving in a narrow channel have many features in common with the
glassy behavior of hard spheres in three dimensions.  In this paper we
study the caging behavior of the disks which sets in at characteristic
packing fraction $\phi_d$.  Four-point overlap functions similar to
those studied when investigating dynamical heterogeneities have been
determined from event driven molecular dynamics simulations and the
time dependent dynamical length scale has been extracted from them.
The dynamical length scale increases with time and, on the
equilibration time scale, it is proportional to the static length
scale associated with the zigzag ordering in the system, which grows
rapidly above $\phi_d$.  The structural features responsible for the
onset of caging and the glassy behavior are easy to identify as they
show up in the structure factor, which we have determined exactly from
the transfer matrix approach.
\end{abstract}



\maketitle
\section{Introduction}
\label{Introduction}

\begin{figure}[ht]
  \centering
  \includegraphics[width=\columnwidth]{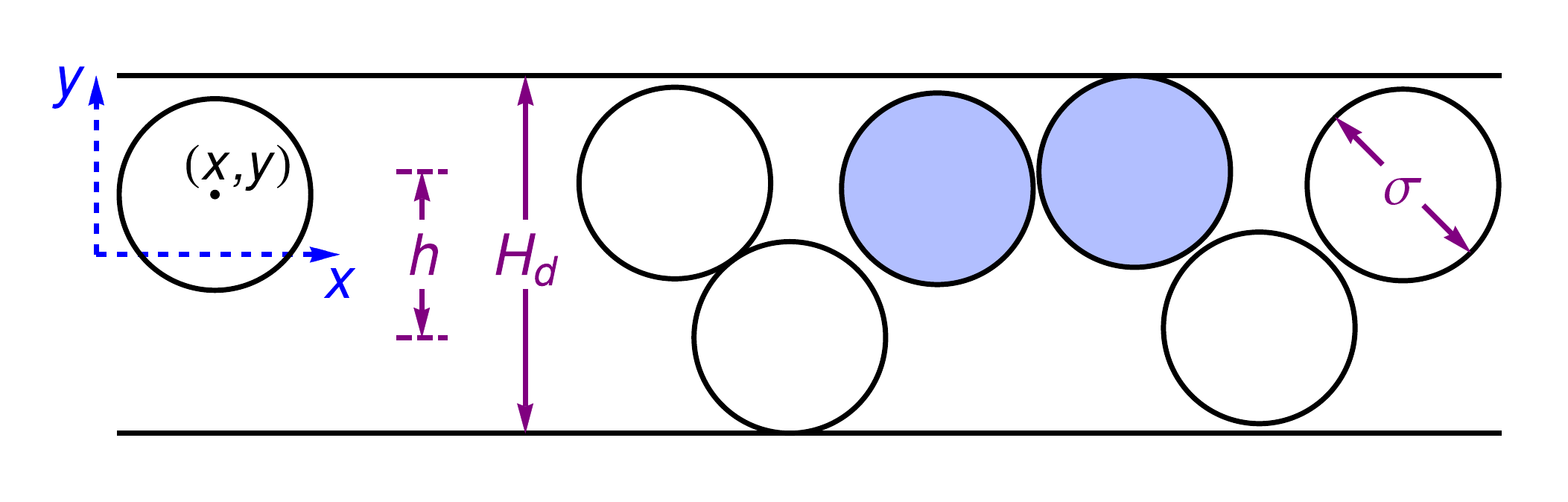} 
  \caption{(Color online) Illustration of the disk--channel system as
    in Ref.~\cite{Godfrey2}.  The shaded disks represent a defect in
    the developing zigzag order.}
  \label{channel}
\end{figure}

It has been frequently observed that disks moving in a narrow channel
can provide useful insights into glassy behavior
\cite{Ivan,Mahdi1,Mahdi2,Mahdi3,AshwinBowles,Godfrey2,Godfrey3}.  In a
recent paper \cite{Godfrey2} two of us studied the static and dynamic
properties of $N$ disks of diameter $\sigma$, which move in a narrow
channel consisting of two impenetrable walls (lines) spaced by a
distance $H_d$, such that $1<H_d/\sigma<1+\sqrt3/2$ (see
Fig.~\ref{channel}).
With channels of this width only nearest-neighbor disks can interact
with one another and neighboring disks cannot pass each other so that
their ordering is preserved: $0\le x_1<x_2<\cdots<x_N\le L$, where
$x_i$ is the position of the center of disk $i$, measured along the
channel, and $L$ is the total length available to the disk centers.
The coordinate $y_i$ of the $i$th disk is measured from the center of
the channel and can vary between $\pm h/2$, where $h=H_d-\sigma$. The
packing fraction $\phi$ in our system of disks is defined as
\begin{equation}
  \phi=\frac{ N \pi \sigma^2}{4 H_d L}\,.
\label{phidef}
\end{equation}
Our disks move ballistically, colliding elastically with each other
and the channel walls.  It is useful for many purposes to regard the
disks as being contained within a system of average length $L$ by
pistons at the ends of the channel which exert a force $f$ to
counteract the momentum transferred by the disks colliding with them.
For any finite value of this force, large fluctuations in the
$x$-coordinates of the disks cause the time-averaged density of disks
to be independent of $x$, so that the system is never crystalline.
The static, equilibrium properties of this simple system can be
determined exactly by use of the transfer matrix
\cite{Godfrey2,Barker,Kofke,VargaBalloGurin,GurinVarga}, but the chief
purpose of this paper to discuss the dynamics.  The dynamical
properties of our system must be determined from simulations, and to
this end we have used event-driven molecular dynamics to handle the
collisions of the disks with each other and with the channel walls.
We find interesting similarities and also some instructive
\emph{differences} with the dynamics of three dimensional
glass-forming liquids.

\begin{figure}[ht]
  \includegraphics[width = \columnwidth]{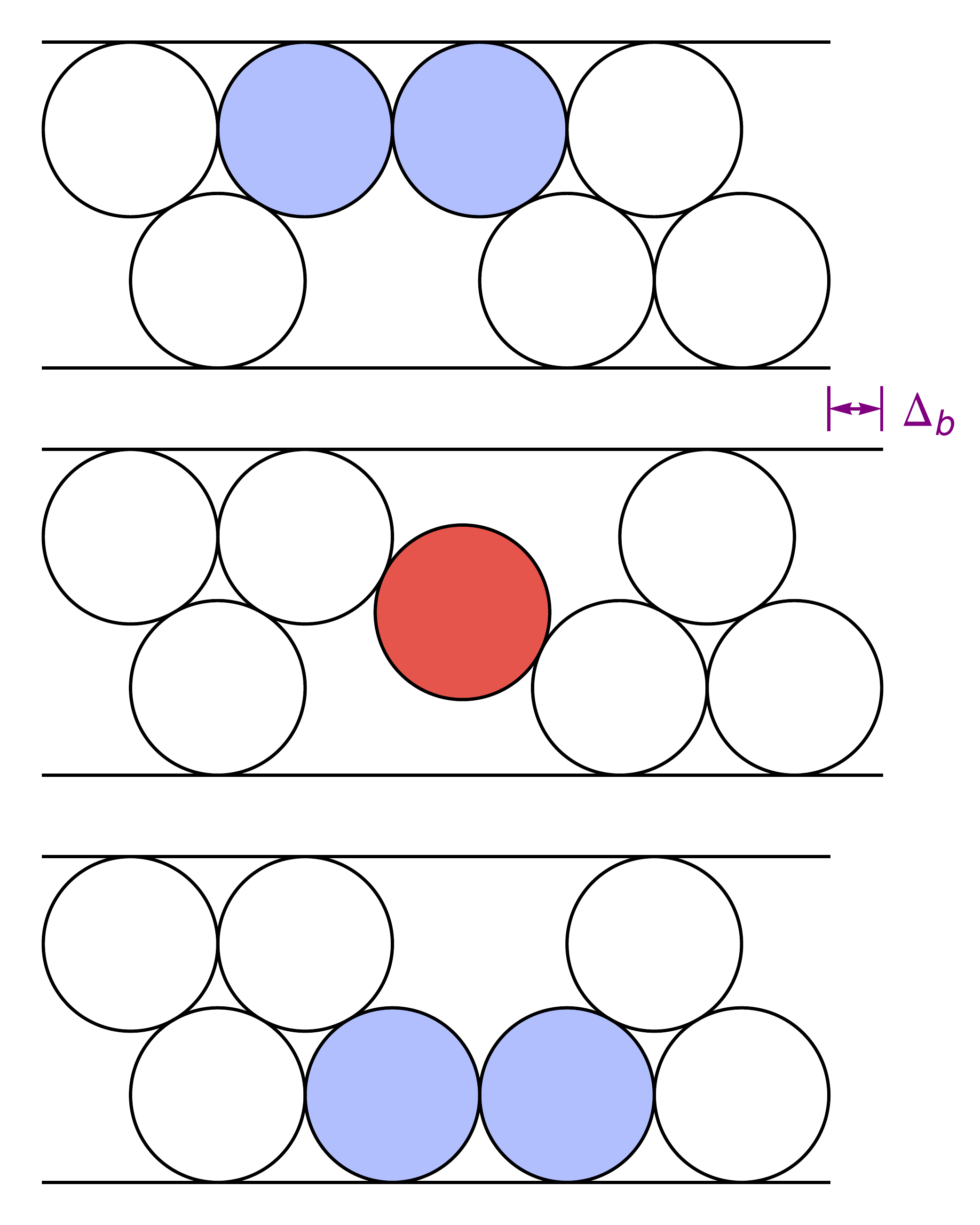} 
  \caption{(Color online) The transition state for motion of a defect.
    In the top and bottom diagrams, the two blue-shaded disks are a
    defect in the zigzag arrangement of disks.  The defect can move
    when one disk crosses the channel by squeezing between its
    neighbors: the system passes through the transition state shown in
    the middle diagram to reach the defect state shown in the bottom
    diagram.  In the top diagram the defect involves disks $3$ and
    $4$; in the bottom diagram the defect involves disks $4$ and $5$,
    when the disks are numbered from the left.  The net motion of the
    defect is to the right, and $\Delta_b$ is the extra length needed
    to allow this motion.}
\label{transitionstate}
\end{figure}

It was found some time ago that in the system of disks in a channel
there is a packing fraction $\phi_d \approx 0.48$ above which the
dynamics are activated \cite{Ivan}.  Similarly, for hard spheres in
three dimensions, the relaxation time grows rapidly above a packing
fraction of $\approx 0.58$.  For a review of this and other features
of hard sphere systems see Ref.~\cite{Zamponi}.  We have also observed
that for $\phi >\phi_d \approx 0.48$ zigzag ordering of the disks
starts to grow rapidly \cite{Godfrey2}; that is, the onset of the slow
dynamics is connected with the growth of this particular kind of
order.  There is a long tradition of associating glassy behavior with
geometrical features associated with the arrangements of particles
around a given particle, see
e.g.~\cite{Royall,Royall0,Royall2,Tarjus,Liu}, and our work on disks
in a narrow channel is entirely consistent with these ideas.  In our
earlier work \cite{Godfrey2,Godfrey3} a length scale $\xi$
associated with zigzag order has been determined from the decay with
$s$ of the correlation function
\begin{equation}
  \langle y_i\, y_{i+s}\rangle\sim (-1)^s \exp (-s/\xi)\,.
\label{xidef}
\end{equation}
It is a measure of the number of disks over which the zigzag order, a
form of bond-orientational order, persists; it is of the same order as
the separation of defects in the zigzag order, where the defects
are of the kind illustrated in the top and bottom panels of
Fig.~\ref{transitionstate}.
In three dimensional systems the ordering associated with glassy
behavior is complicated \cite{Liu}.  Furthermore, the ordering
associated with glassy behavior is not apparent in changes to the
structure factor as it is cooled through the glass transition.  In
Sec.~\ref{structurefactor} we calculate the structure factor of our
system, defined as
\begin{equation}
  S(k_x, k_y)=
  \frac{1}{N} \sum_{i,j}\langle \exp[i k_x (x_i-x_j)+i k_y(y_i-y_j)]\rangle,
\label{structuredef}
\end{equation}
at packing fractions close to $\phi_d$ by use of the transfer matrix
formalism and so it is exact to numerical accuracy.  It is found to
change rapidly near $\phi_d$.  This marked difference with the
behavior in three dimensions where the structure factor hardly alters
near the glass transition indicates that glass behavior in three
dimensions must, as suggested in
Refs.~\cite{Royall,Royall0,Royall2,Tarjus,Liu}, involve higher order
correlations which have little impact on the structure factor, unlike
the simple zigzag orientational order of the narrow channel system.

In Sec.~\ref{plateau} we shall study the onset of the slow dynamics
which sets in around $\phi_d$.  In three dimensions the slow dynamics
are normally attributed to the onset of caging behavior where a
particle is trapped by its neighbors and this behavior is believed to
be captured by the mode-coupling approach.  The particle can escape
its cage on the alpha relaxation time $\tau_{\alpha}$.  To determine
whether or not caging occurs in our system we have studied the
mean-square displacements
\begin{equation}
  \Delta^2(t)=
  \big\langle \frac{1}{N} \sum_{i=1}^N|y_i(t)-y_i(0)|^2\big\rangle\,,
\label{Deltadef}
\end{equation}
where the average is over different initial states.  We find that
there is caging of particles in our system, i.e. there is (at large
enough packing fractions) a plateau in $\Delta^2(t)$ before its long
time limit is reached.  We find that in our system there are
\emph{two} distinct large time scales, which we call $\tau$ and
$\tau_D$ following Ref.~\cite{Godfrey2}; we can obtain both of them
from the behavior of $\Delta^2(t)$.  The smaller of these time scales,
$\tau$, is the typical time scale for a disk to cross from one side of
the system to the other by the process shown in
Fig.~\ref{transitionstate}.  It marks the time at which the particle
starts to escape from its cage or the end of the plateau in
$\Delta^2(t)$: in three dimensional systems this would be called the
alpha relaxation time~$\tau_{\alpha}$.  At packing fractions above
$\phi_d$, escape from a cage can be regarded as a displacement of a
defect in the zigzag ordering of the disks as is also illustrated in
Fig.~\ref{transitionstate}.  We are able to determine from the
simulation the diffusion constant for the movement of such defects.
Its dependence on packing fraction is consistent with the general
picture of a crossover from fragile glass behavior at low packing
fractions to strong glass behavior at high packing fraction, which has
been investigated previously in Refs.~\cite{Mahdi1,Mahdi3}.

The second long time scale $\tau_D$ is essentially the longest time
scale in the system.  In equilibrium defects are thermally nucleated
in pairs and the defects produced then diffuse and annihilate with
each other.  It is this process of diffusion with creation and
annihilation which takes place on the time scale $\tau_D$.  There is a
simple relation between $\tau$ and $\tau_D$: $\tau_D \sim \tau \xi^2$
\cite{Godfrey2}.  Note that $\xi$ is the typical distance between
defects in equilibrium, so that $\xi^2/D$ is the time it would take
for a defect to move a distance $\xi$: this gives a diffusion
coefficient $D$ for the defects which varies as
$1/\tau$~\cite{Godfrey2}.  $\tau_D$ is determined from the time at
which $\Delta^2(t)$ approaches its equilibrium limit.

A characteristic of glassy dynamics is the appearance of a plateau in
the decay of certain time-dependent correlation functions.  This
plateau eventually decays to zero after the time $\tau_{\alpha}$, the
alpha relaxation time.  The time to reach the plateau is the beta
relaxation time $\tau_{\beta}$.  Correlation functions with this
feature have not previously been studied in our narrow channel system.
The existence of a plateau in $\Delta^2(t)$ implies that there will be
such a plateau in the decay of
\begin{equation}
  R(t)=\frac{1}{N}\sum_i \langle y_i(0)\,y_i(t) \rangle\,;
\label{Qdef1}
\end{equation}
at long times $R(t)$ approaches zero.  As noted above, our results
support a model in which the diffusion of defects leads to
equilibrium.  In models of this kind, $R(t)$ is expected to decay with
a \emph{stretched} exponential form $\exp(-[t/\tau_D]^{1/2})$, where
$\tau_D$ is the equilibration time of the system~\cite{Redner}.

We have also studied a correlation function related to the $\chi_4(t)$
which has been much studied in three
dimensions~\cite{Glotzer,Berthier,Royall2}.  In three dimensions
$\chi_4(t)$ rises to a peak at the alpha relaxation time and then
decays to zero as the particles escape their cages.  Our $\chi_4(t)$
increases to a constant, plateau value as $t$ increases and stays at
this value as $t \to \infty$.  In this regard the narrow channel
system behaves more like a spin glass than a structural glass
\cite{BB05}.  We have also extracted a \emph{dynamical} correlation
length $\xi_4(t)$ from a four-point correlation function.  In our
system as $t \to \tau_D$, $\xi_4$ grows towards the static correlation
length $\xi$, which measures the extent of zigzag, i.e. structural,
order.

In Sec.~\ref{structurefactor} we show how the structure factor can be
determined exactly from the transfer matrix procedure and we
demonstrate that it changes rapidly for densities close to $\phi_d$.
Our dynamical studies are in Sec.~\ref{plateau}.  We conclude with a
discussion in Sec.~\ref{conclusions}.

\begin{figure}[ht]
  \includegraphics[width = \columnwidth]{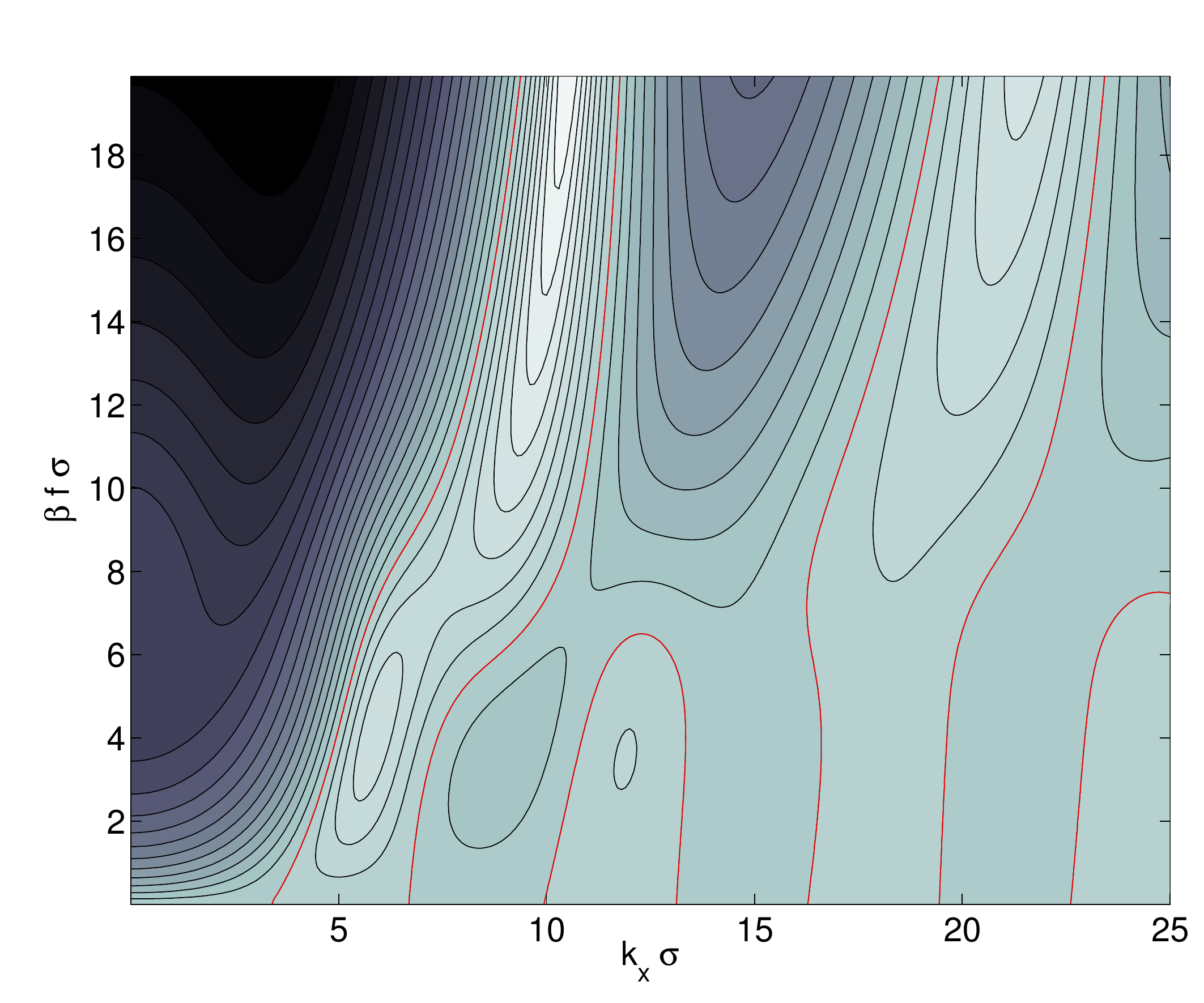}
  \caption{(Color online) The structure factor $S(k_x,k_y=0)$ as a
    function of $\beta f\sigma$ and~$k_x\sigma$ for the narrow-channel
    system with $h=\sqrt3\,\sigma/2$.  The position of the first peak
    in $S$ (near $k_x\sigma=6$ for $\beta f\sigma=4$) changes rapidly
    for $\beta f\sigma$ in the range 6.5 to~8, following the change in
    periodicity that accompanies the growth of zigzag order.  Adjacent
    contours are spaced by $0.1$ in $\log_{10} S$ and contour lines
    with $S=1$ are marked in red.}
\label{str-fac-ky-zero}
\end{figure}

\begin{figure}[ht]
  \includegraphics[width = \columnwidth]{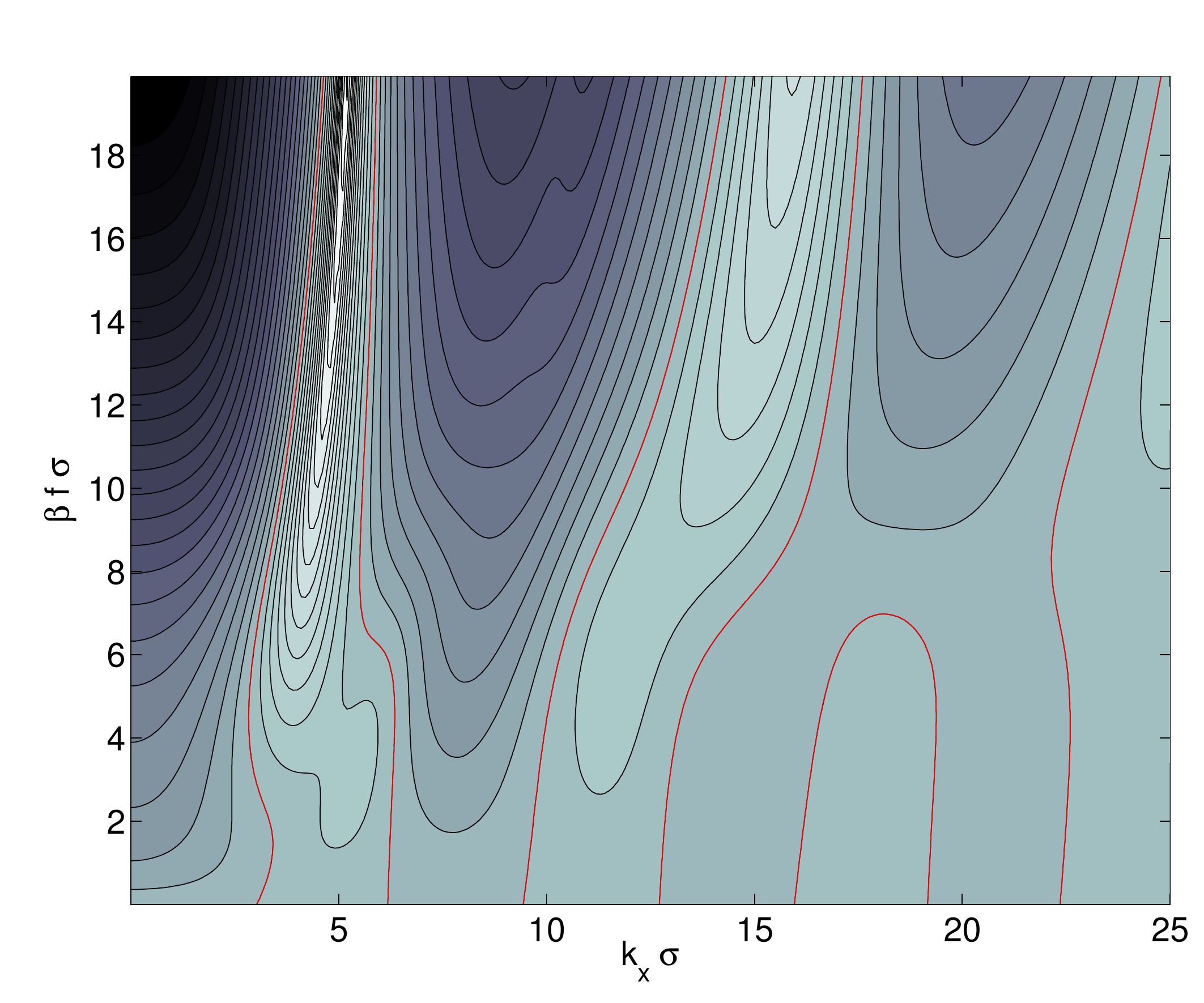}
  \caption{(Color online) The structure factor $S(k_x,k_y=\pi/h)$ as a
    function of $\beta F\sigma$ and~$k_x\sigma$ for the narrow-channel
    system with $h=\sqrt3\,\sigma/2$.  A peak near $k_x\sigma=4$ grows
    rapidly for $\beta f\sigma>4$, consistent with the growth of
    zigzag correlations.  Contour lines are described in the caption
    to Fig.~\ref{str-fac-ky-zero}.}
\label{str-fac-ky-pi}
\end{figure}

\section{Structure Factor}
\label{structurefactor}

The structure factor is well known from the part it plays in the
scattering of electromagnetic radiation by liquids~\cite{Frenkel}.  It
is the Fourier transform of the density--density correlation function
and so provides information on the relative positions of scatterers
within the system.  Like a liquid, our system of disks in a channel
has no long-range order (along the channel) for finite values of the
longitudinal force~$f$; but, unlike in the bulk of a liquid, the
short-range order is strongly affected by the presence of confining
walls, leading to the zigzag correlations discussed in
Sec.~\ref{Introduction}.  In this section we show that the structure
factor can be calculated essentially exactly for disks in a narrow
channel.  Our numerical results for the case $h=\sqrt3\sigma/2$ show a
rapid change in the short-range order for $\beta f\sigma$ in the range
$6.5$ to~$8$, which corresponds to packing fractions $\phi$ in the
range $0.45$ to~$0.50$.  This range of packing fractions correlates
closely with the onset of activated dynamics, as we discuss later in
Sec.~\ref{plateau}.

In the limit $N\to\infty$, the definition of the structure factor,
Eq.~\eqref{structuredef}, may be rewritten in the form
\begin{equation}
  S(k_x,k_y)
  = \sum_{n=-\infty}^\infty S_n
  = 1 + 2\Re\sum_{n=1}^\infty S_n\,,
\label{SFsum}
\end{equation}
where
\begin{equation}
  S_n=\langle e^{i(k_x[x_n-x_0] + k_y[y_n-y_0])}\rangle\,.
\label{Sndef}
\end{equation}
At $k_x=0$, $S(k_x,k_y)$ has a delta-function singularity and the sum
on the right-hand side of \eqref{SFsum} diverges.  As we now show,
for $k_x\ne0$ the sum can be evaluated relatively simply by solving a
pair of integral equations.  One of these equations is known from the
transfer-matrix formalism introduced by Barker~\cite{Barker} and
applied by Kofke and Post \cite{Kofke} to the problem of hard disks in
a channel.  We follow the latter authors in using an ensemble in which
the longitudinal force $f$ is constant and we refer the reader to
their paper \cite{Kofke} for details.

Let $\psi_n(y)$ be the eigenfunctions of Kofke and Post's integral
equation
\begin{equation}
  \lambda_n\,\psi_n(y_1) =
  \int_{-h/2}^{\,h/2} e^{-\beta f\sigma_{1,0}}\,\psi_n(y_0)\,dy_0\,,
\label{integraleqKP}
\end{equation}
where $y_0$ and $y_1$ are the $y$-coordinates of a neighboring pair of
disks and $\sigma_{1,0}=[\sigma^2 - (y_1-y_0)^2]^{1/2}$ is the
distance of closest approach of their centers, measured along the
$x$-axis.  Approximations to the eigenfunctions $\psi_n$ and the
eigenvalues $\lambda_n$ can be found by discretizing
Eq.~\eqref{integraleqKP}, which converts it to a real-symmetric matrix
eigenvalue problem.  The eigenfunctions (taken to be real) can be
normalized so that
\begin{equation}
  \int [\psi_n(y_1)]^2\,dy_1 = 1\,.
\label{psinorm}
\end{equation}
In this and subsequent equations, the limits of the $y$ integration are
$-h/2$ and $h/2$, the same as in Eq.~\eqref{integraleqKP}.

Equilibrium expectation values, such as those needed for the
quantities $S_n$ defined in \eqref{Sndef}, can be expressed as
integrals involving the eigenfunction $\psi_1$ which corresponds to
the largest eigenvalue~$\lambda_1$.  We illustrate this for the
calculation of $S_1$ below.

$S_1$ is the expectation value of $\exp(ik_x[x_1-x_0]+ik_y[y_1-y_0])$,
which is a function of $y_0$, $y_1$ and the gap $s$ between disks 0
and~1, defined by
\begin{equation}
  s + \sigma_{1,0} \equiv x_1-x_0\,.
\end{equation}
From the results of Ref.~\cite{Kofke}, the probability distribution
for the variables $y_0$, $y_1$, and $s$ is proportional to
\begin{equation*}
  \psi_1(y_1)\,e^{-\beta f[s+\sigma_{1,0}]}\,\psi_1(y_0)\,.
\end{equation*}
\begin{widetext}\noindent
Accordingly, $S_1$ is given by
\begin{equation}
  S_1 =
  \langle e^{ik_x(s+\sigma_{1,0}) + ik_y(y_1-y_0)}\rangle =
  \frac{\int \psi_1(y_1)\int\int_0^\infty
    e^{(ik_x-\beta f)(s+\sigma_{1,0}) +ik_y(y_1-y_0)}\,\psi_1(y_0)\,ds\,dy_0\,dy_1}
       {\int \psi_1(y_1)\int\int_0^\infty
         e^{-\beta f(s+\sigma_{1,0})}\,\psi_1(y_0)\,ds\,dy_0\,dy_1}\,.
\end{equation}
After completing the integrals with respect to $s$ and using the
eigenvalue equation \eqref{integraleqKP} and the normalization
condition \eqref{psinorm} to simplify the denominator, we obtain
\begin{align}
  S_1 &=
  \int \psi_1(y_1)
  \left\{
  \frac{\beta f}{\lambda_1(\beta f - ik_x)}
  \int
  e^{(ik_x-\beta f)\sigma_{1,0} +ik_y(y_1-y_0)}\,\psi_1(y_0) \,dy_0
  \right\}
  dy_1\nonumber\\
  &\equiv
  \int \psi_1\,\hat S \psi_1\, dy_1\,,
\end{align}
\end{widetext}
in which the bracketed expression in the first line defines the action
of the integral operator $\hat S$ on~$\psi_1$.  More generally, for
$n\ge1$ one can write
\begin{equation}
  S_n = \int \psi_1\,\hat S^n\psi_1\,dy_1\,,
\end{equation}
so that the sum in Eq.~\eqref{SFsum} becomes
\begin{equation}
  \sum_{n=1}^\infty S_n
  = \sum_{n=1}^\infty\int \psi_1\,\hat S^n \psi_1\,dy_1
  = \int \psi_1\,\hat S\phi\,dy_1\,,
\end{equation}
where $\phi$ is the solution of
\begin{equation}
  \phi = \psi_1 + \hat S \phi\,,
\label{integraleqphi}
\end{equation}
which is a Fredholm equation of the second kind.  Given the function
$\psi_1$ found by solving the discretized Eq.~\eqref{integraleqKP},
the calculation of $\phi$ requires only the solution of the set of
linear equations obtained by
discretizing~Eq.~\eqref{integraleqphi}.

Finally, in terms of $\psi_1(y_1)$ and~$\phi(y_1)$, the structure
factor is given by
\begin{equation}
  S(k_x,k_y) = 1 + 2\Re\int\psi_1\,\hat S\phi\,dy_1 \,,
\end{equation}
which depends on $k_x$ and $k_y$ via $\phi$ and the operator~$\hat S$.

As indicated above, the equations \eqref{integraleqKP} and
\eqref{integraleqphi} can be solved by discretization.  For all but
the smallest values of $\beta f\sigma$, the function $\psi_1$ is
concentrated near the walls and it is sampled at smaller intervals in
those regions, to keep the dimension of the matrices relatively small.
To implement this nonuniform sampling we make the change of variable
\begin{equation}
  u(y) = \frac{\sinh(\alpha\beta f y)}{\sinh(\alpha\beta f h/2)}\,,
\label{uofy}
\end{equation}
where $\alpha = \tfrac12h/\sqrt{\sigma^2-h^2}$ and the values of $u$
are taken to be uniformly spaced in the interval $[-1,1]$.  The
symmetry of the transfer matrix is preserved by solving
Eq.~\eqref{integraleqKP} for the function $[dy/du]^{1/2}\psi_1$,
rather than $\psi_1$.  Matrices of dimension $100\times100$ are
sufficient to reproduce the results presented in this Section.

\begin{figure}[ht]
  \includegraphics[width = \columnwidth]{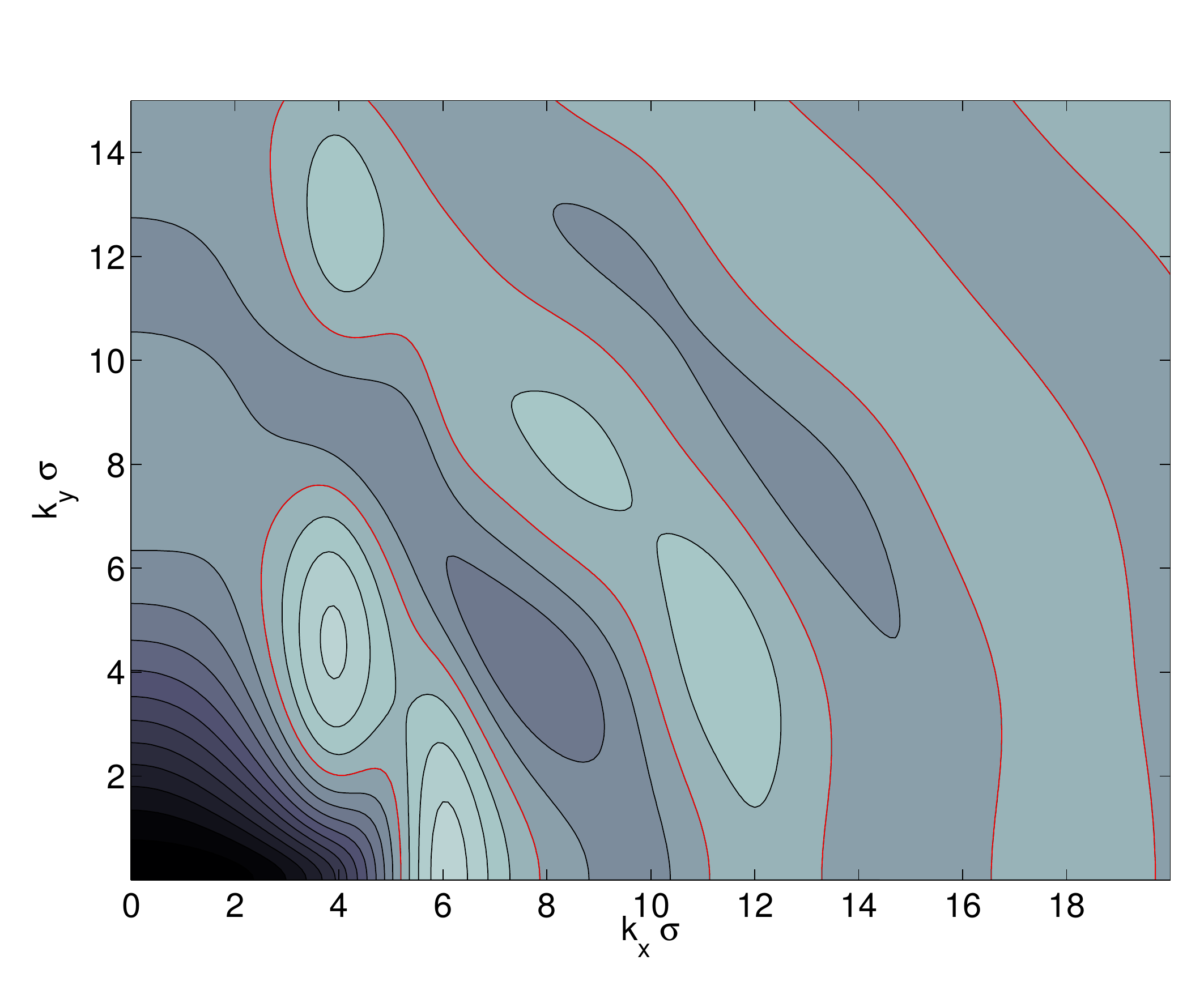}
  \caption{(Color online) The structure factor $S(k_x,k_y)$ for $\beta
    f\sigma=5$ for the narrow-channel system with
    $h=\sqrt3\,\sigma/2$.  Contour lines are described in the caption
    to Fig.~\ref{str-fac-ky-zero}.}
\label{str-fac-f-5}
\end{figure}

\begin{figure}[ht]
  \includegraphics[width = \columnwidth]{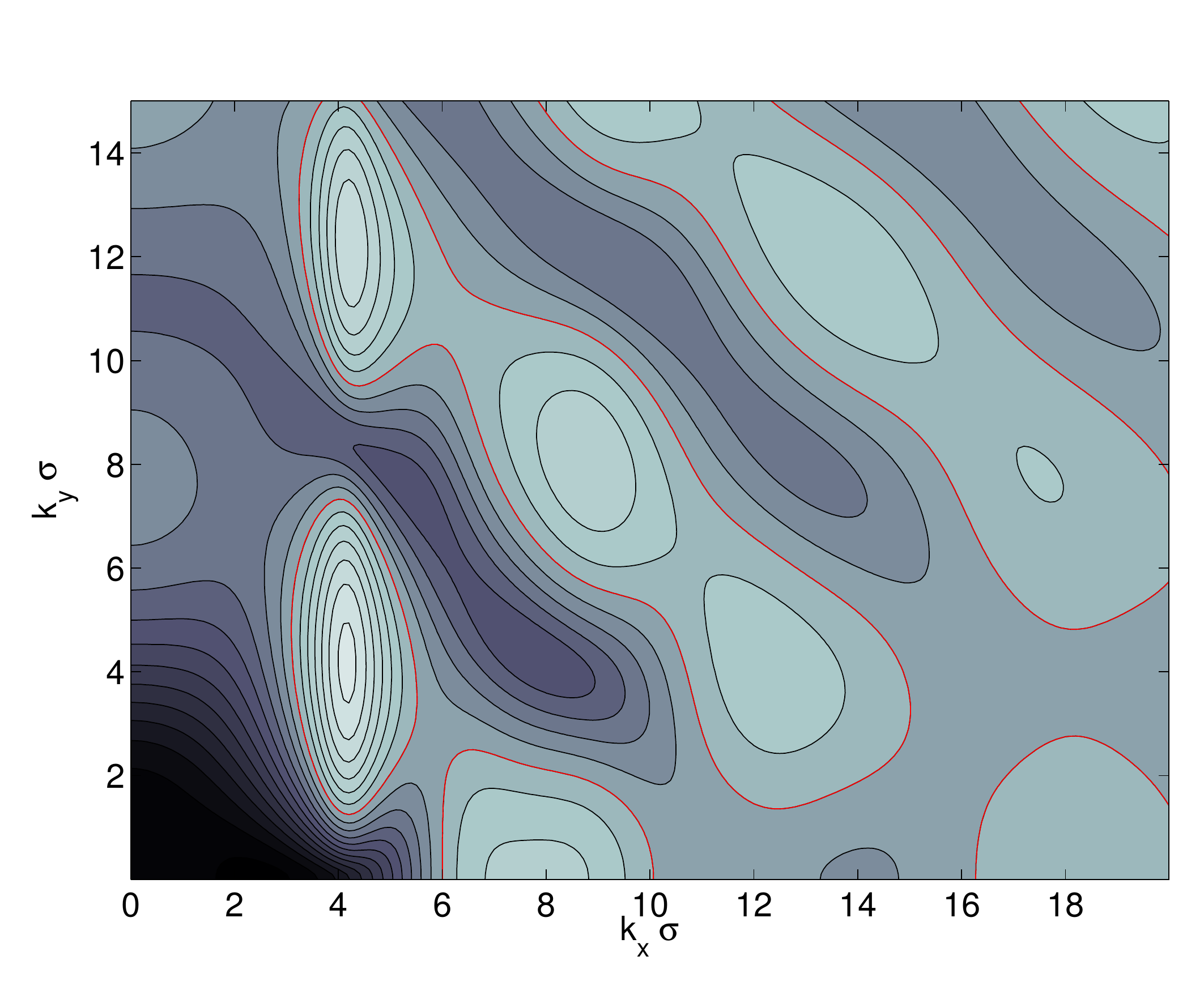}
  \caption{(Color online) The structure factor $S(k_x,k_y)$ for $\beta
    f\sigma=7.5$ for the narrow-channel system with
    $h=\sqrt3\,\sigma/2$.  Contour lines are described in the caption
    to Fig.~\ref{str-fac-ky-zero}.}
\label{str-fac-f-7_5}
\end{figure}

\begin{figure}[ht]
  \includegraphics[width = \columnwidth]{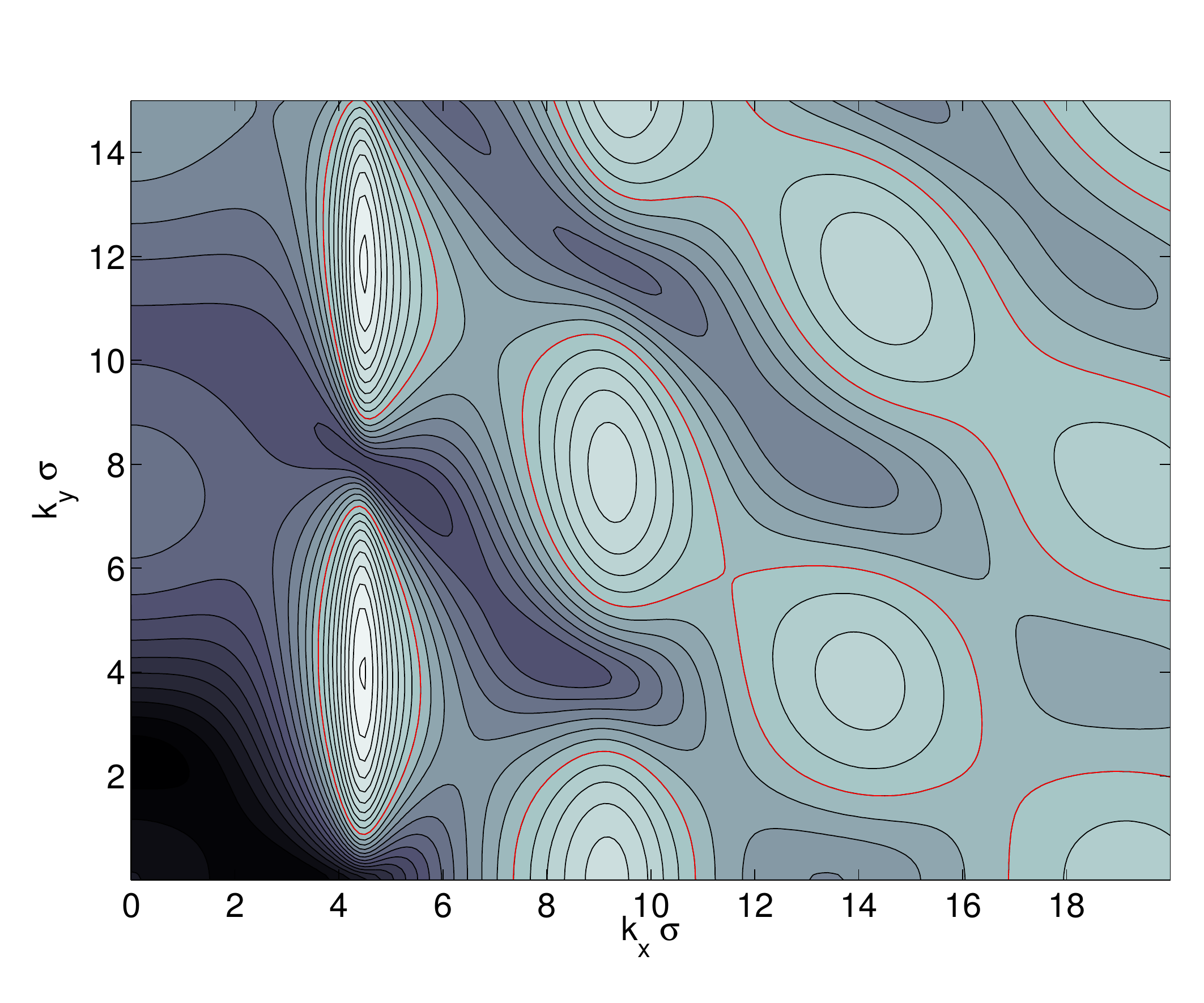}
  \caption{(Color online) The structure factor $S(k_x,k_y)$ for $\beta
    f\sigma=10$ for the narrow-channel system with
    $h=\sqrt3\,\sigma/2$.  Contour lines are described in the caption
    to Fig.~\ref{str-fac-ky-zero}.}
\label{str-fac-f-10}
\end{figure}

Numerical results are shown in
Figs.~\ref{str-fac-ky-zero}--\ref{str-fac-f-10}.  In
Fig.~\ref{str-fac-ky-zero}, the structure factor is plotted as a
function of $\beta f\sigma$ and $k_x$ for the case $k_y=0$.  For
$k_y=0$, the structure factor is sensitive only to correlations in the
$y$-averaged density.  Zigzag order is growing rapidly for $\beta
f\sigma$ in the range 6.5 to~8, and this can be seen in
Fig.~\ref{str-fac-ky-zero} as a rapid change in the nature of the
first maximum with respect to $k_x$.  For very small values of $f$,
the oscillations in $S(k_x,k_y=0)$ are small and (much as for a
low-density gas of hard rods) their presence is due to the
discontinuity of the pair distribution function $g(r)$ at $r=\sigma$.
But as $f$ increases, a new peak emerges and rapidly becomes dominant.
Its position reflects the changing spatial periodicity in the
$x$-direction: for $\beta f\sigma\to\infty$, the peak approaches
$k_x=4\pi/\sigma$, which can be understood as $2\pi/(\sigma/2)$, where
$\sigma/2$ is the spatial period of the $y$-averaged density in this
limit.

Figure~\ref{str-fac-ky-pi} shows the evolution of the structure factor
for the case $k_y=\pi/h$, where the value of $k_y$ has been chosen to
better illustrate the growth of zigzag order.  Note that when $\beta
f\sigma$ is large, nearest-neighbor disks are separated by $\Delta
x\approx\sigma/2$ along the channel and by $\Delta y\approx\pm h$ in
the transverse direction.  For $k_y=\pi/h$, the nearest neighbors will
scatter in phase when $k_x\,\Delta x\approx \pi$, i.e.\ for
$k_x\approx2\pi/\sigma$.  This scattering results in a peak in the
structure factor that is clearly visible in Fig.~\ref{str-fac-ky-pi}.
The peak first appears near $k_x\sigma=4$ for $\beta f\sigma \approx
4$.  It strengthens as $\beta f\sigma$ increases and it eventually
approaches $k_x\sigma=2\pi$.

Figures \ref{str-fac-f-5}--\ref{str-fac-f-10} show the structure
factor for three values of the force for which $S(k_x,k_y)$ is
changing most rapidly: $\beta f\sigma=5$, $7.5$, and~$10$.  The maxima
in $S$ grow and become narrower (in the $k_x$ direction) as $f$
increases and the zigzag correlations strengthen.  For very large $f$,
the widths of the peaks decrease as $(\beta f\sigma)^{-2}$.  We note
that this $f$-dependence of the peak-width is the same as is found for
a one-dimensional gas of hard rods, whose structure factor was derived
analytically by Zernike and Prins~\cite{ZernikePrins}.

\section{Time-dependent behavior}
\label{plateau}

In order to study time dependent effects we have used event driven
molecular dynamics based upon the code referred to in
Ref.~\cite{Skoge}; the speed of the program was improved by using the
fact that in our narrow-channel system only nearest-neighbor disks can
collide.  The initial state from which the system evolves was created
by means of the Lubachevsky--Stillinger algorithm \cite{LS}, starting
from a random configuration of small disks.  Their diameters were
slowly increased to the desired value during the course of a
simulation which preceded the long runs used to study dynamics in the
equilibrated system.

The force $f$ along the channel was computed by using a virial
formula.  Suppose that at the instant of collision between two disks,
the $x$-separation of their centers is $\Delta x(c) >0$, where $c$
labels the collision.  The $x$-component of the momentum transferred
from the disk at smaller $x$ to the disk at larger $x$ is a positive
quantity, $\Delta p_x(c)>0$.  With the values of $\Delta x(c)$ and
$\Delta p_x(c)$ determined by simulation, the average longitudinal
force can be found from
\begin{equation}
  f L =
  \frac N\beta +
  \frac1{\tau_{\rm sim}} \sum_c \Delta x(c)\,\Delta p_x(c)\,,
\end{equation}
where the sum includes all disk--disk collisions $c$ that occur during
the simulation time~$\tau_{\rm sim}$.

\begin{figure}[ht]
  \centering
  \includegraphics[width=\columnwidth]{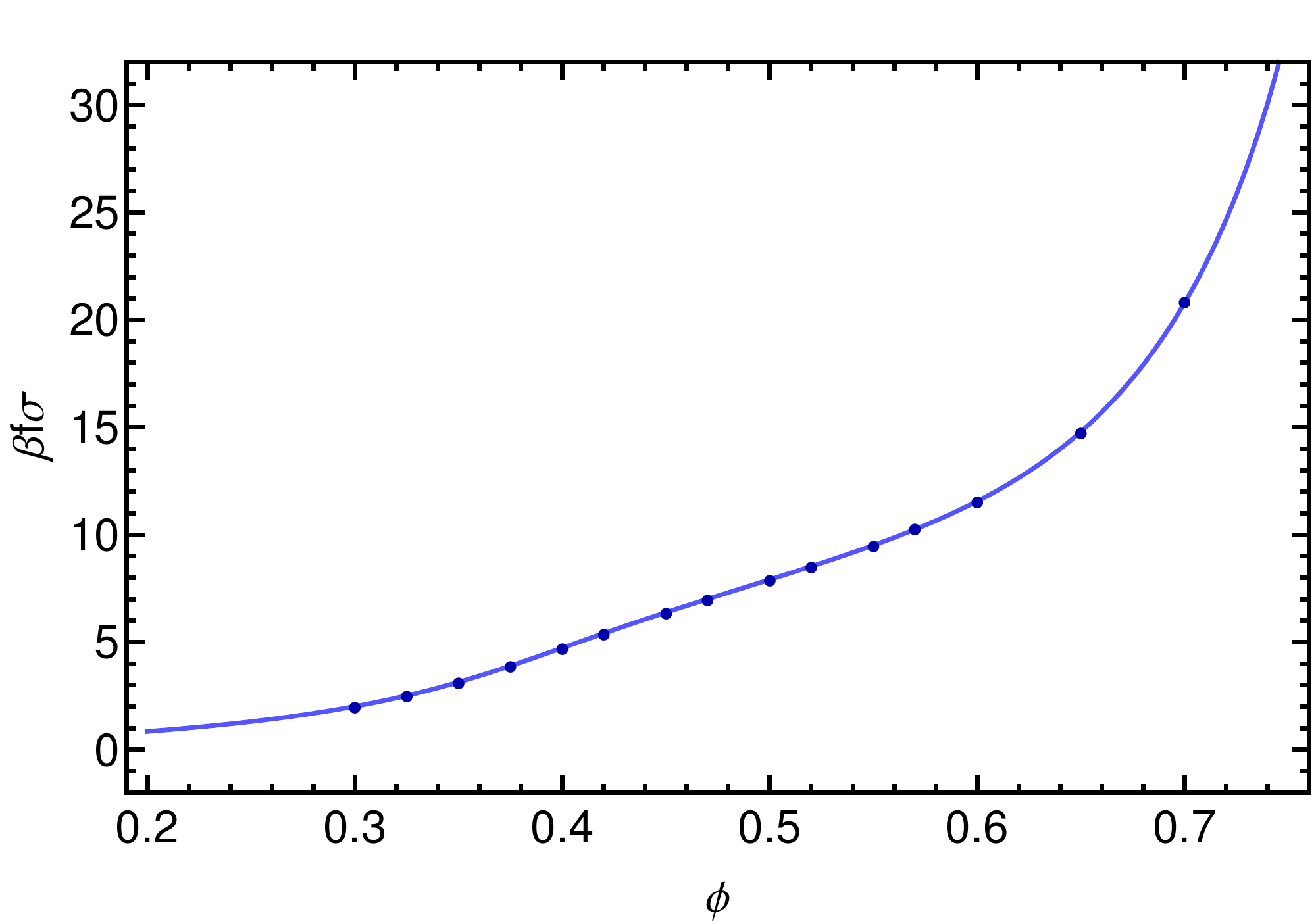} 
  \caption{(Color online) Equation of state $\beta f \sigma$ versus
    $\phi$ for the narrow-channel system with $h=\sqrt3\sigma/2$.  The
    solid line shows the equation of state calculated from the
    transfer matrix results of Ref.~\cite{Godfrey2}, while the data
    points were obtained from long-time simulations of $N=10000$
    disks.}
  \label{bfs}
\end{figure}

All of our simulations were performed with $N =10000$, $h=\sqrt3/2$,
and $\beta=\sigma=m=1$, where $m$ is the mass of a disk.  To check
their accuracy we calculated the equation of state and compared it
with the results of the exact transfer matrix calculation
\cite{Godfrey2}, as shown in Fig.~\ref{bfs}.
The agreement is excellent up to $\phi =0.65$.  There is a discrepancy
at $\phi=0.70$ (but too small to be visible in the figure) as at this
packing fraction the time scale for equilibration, $\tau_D$, is
becoming comparable to our simulation time.

\subsection{Time scales}
\label{timescales}

\begin{figure}[ht]
  \centering
  \includegraphics[width=\columnwidth]{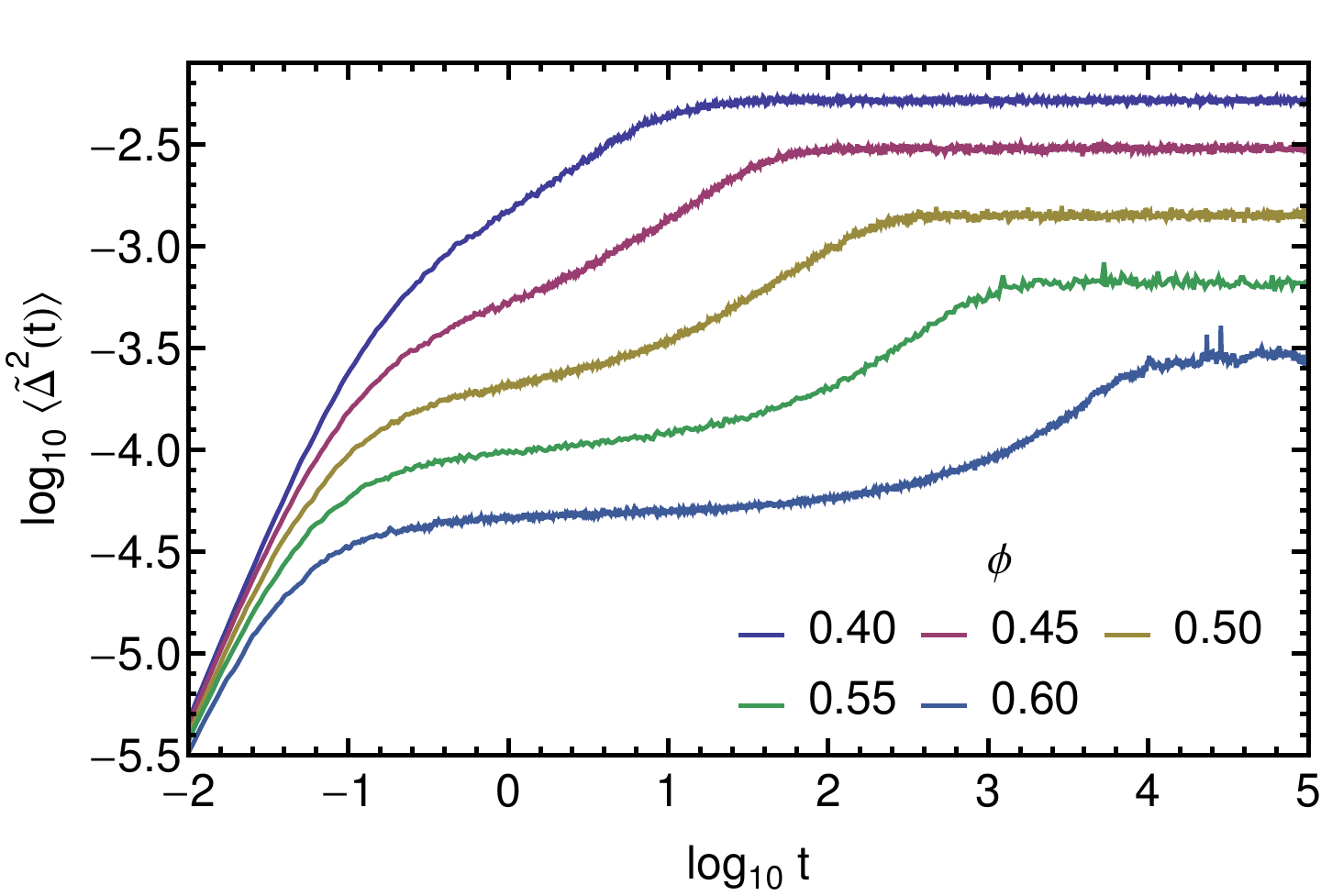} 
  \caption{(Color online) Time dependence of the mean-squared
    displacements (in units of $\sigma^2$), of 10000 disks defined via
    Eq.~\eqref{rattlerout}, a quantity which was introduced in
    Ref.~\cite{ikeda:rattlers} to minimize the contribution from the
    most mobile disks.  The results for the various packing fractions
    $\phi$ which were studied are averaged over 20 independent
    equilibrated trajectories.  Time is in units of $(\beta m
    \sigma^2)^{1/2}$.}
  \label{delta_squ}
\end{figure}

To see the emergence of caging behavior it is convenient to study
\begin{equation}
  \tilde{\Delta}^2(t)=
  \frac{1}{N}\sum_{i=1}^{N}\frac{1}{\langle|y_i(t)-y_i(0)|^{-2}\rangle},
\label{rattlerout}
\end{equation}
where the angled brackets indicate an average over runs with different
initial states.  Though it appears unnatural at first sight, this
quantity was introduced in Ref.~\cite{ikeda:rattlers} to miminize the
contribution of rattlers.  The ability of rattlers to move a distance
of $\mathrm{O}(\sigma)$ tends to dominate the mean-squared
displacements at small times $t$.  We have a similar problem here in
that a few disks which border gaps are able to cross from one side of
the channel without hindrance.  In Fig.~\ref{delta_squ} one can see
the emergence of a plateau as $\phi$ increases from $0.45$ to $0.50$,
which suggests that caging of the disks is setting in as the zigzag
order develops around~$\phi_d$.  It was in Ref.~\cite{Ivan} that
$\phi_d \approx 0.48$ was first identified as the onset point for
activated dynamics.

\begin{figure}[ht]
  \centering
  \includegraphics[width=\columnwidth]{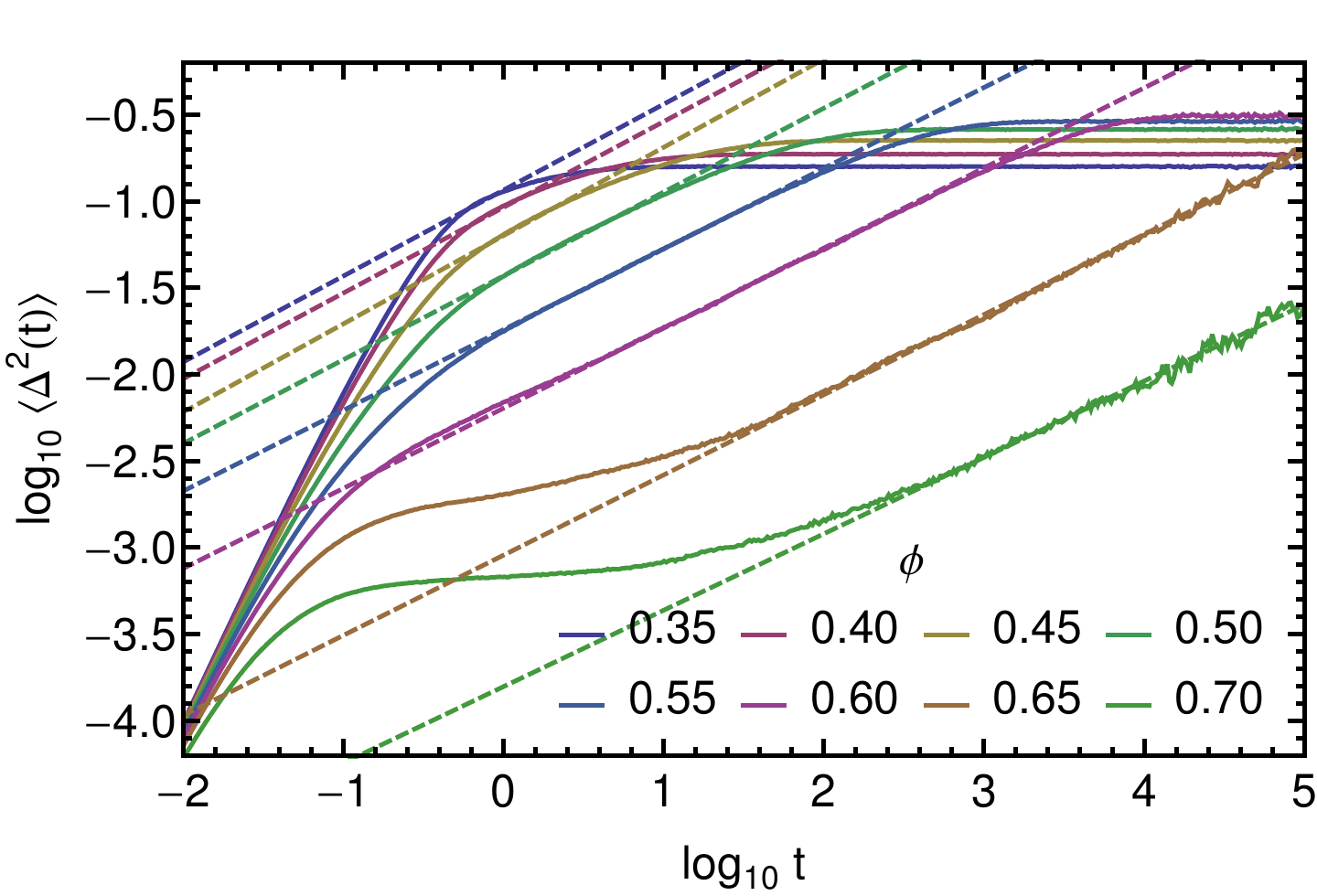} 
  \caption{(Color online) Time dependence of the mean-squared
    displacements (in units of $\sigma^2$), for trajectories of 10000
    disks at varying packing fractions $\phi$.  Time is in units of
    $(\beta m\sigma^2)^{1/2}$.  The dashed lines fit the central
    regions where the gradient is approximately $0.5$, indicating a
    process of relaxation that is dominated by the diffusion of
    defects, as suggested in \cite{Godfrey2} as a mechanism for the
    $\alpha$-relaxation.  For $\phi \gtrsim 0.60$ a plateau is seen to
    form between the small-$t$ ballistic and large-$t$ diffusive
    regimes.  The plateau corresponds to disks becoming trapped in
    cages formed by their neighboring disks, which require cooperative
    motion to break.  At very long times $\Delta^2(t)$ tends to a
    constant value, $2\langle y_i^2\rangle$.}
  \label{dy_squ}
\end{figure}

It is easier to understand the behavior of the unmodified mean-square
displacement $\Delta^2(t)$ as defined in Eq.~\eqref{Deltadef}.
However, Fig.~\ref{dy_squ} shows that for this quantity the plateau is
only clearly visible at values of $\phi$ above $0.60$, which is well
into the density regime where the dynamics are
activated~\cite{Ivan,Godfrey2}.

Some of the features on display in Fig.~\ref{dy_squ} include:
\begin{enumerate}
\item Final long-time limit: In the limit $t \to \infty$,
  $\Delta^2(t)$ reaches a finite value.  (In three dimensions in the
  same limit, the mean-square displacement of a particle increases
  without limit linearly with $t$.)  From its definition,
  $\Delta^2=\langle y_i^2(t)+y_i^2(0)-2 y_i(t) y_i(0) \rangle$; the
  last term in the brackets gives $2 R(t)$, which tends to zero in the
  long time limit, so $\Delta^2$ tends to $2 \langle y_i^2\rangle$, a
  quantity which is close to $2 (h/2)^2$ in the limit of large $f$
  when the disks are mostly pushed to the sides of the channel.

\item For small $t$, $\Delta^2(t)$ increases as $t^2$; i.e., the
  motion is ballistic.  This regime is larger at smaller values of
  $\phi$, for which the gaps between disks are larger.

\item A ``shoulder'' begins to form around $\phi \sim 0.60$, and
  develops into a clearly visible plateau for $\phi \gtrsim
  0.65$.  This is a clear analogue of the caging effect seen in three
  dimensions.  It sets in at higher packing fractions than $\phi_d
  \approx 0.48$, the packing fraction at which the growing zigzag
  order results in the dynamics becoming activated.

\item Beyond the ``glassy'' plateau for $\phi \ge 0.60$, and above the
  ballistic regime for $\phi \le 0.60$, there is a time scale $\tau$
  beyond which a power-law dependence on $t$ sets in,
  $\Delta^2(t)\propto t^{1/2}$.  This dependence on $t$ is due to the
  slow diffusion of defects in the zigzag arrangement of disks.
\end{enumerate}

We can explain some of these features in greater detail.  While
activated dynamics may set in around a packing fraction $\phi \approx
0.48$, some disks still find at this density that they can easily
cross the channel, and it is not until a packing fraction of $0.60$
that the numbers of these ``rattling'' disks become negligible.  The
plateau represents a clear caging effect, and it lasts for a time
$\tau$, the time it typically takes for a disk to cross from one side
of the channel to the other by the transition state mechanism depicted
in Fig.~\ref{transitionstate}.  Once the zigzag order sets in, the
motion of the defects as in Fig.~\ref{transitionstate} dominates the
behavior of most of the disks.  However, at packing fractions around
$0.48$ there are still some disks that can travel from one side of the
channel to the other with little hindrance from their neighbors.  (A
similar observation has been made in Ref.~\cite{Mahdi3}.)  Their
contribution to $\tilde{\Delta}^2(t)$, defined in
Eq.~\eqref{rattlerout} is small, which enables one to see the
emergence of the caging behavior in it at lower packing fractions than
for $\Delta^2(t)$.

The time scale $\tau$ for a defect to move as shown in
Fig.~\ref{transitionstate} was studied numerically in Ref.~\cite{Ivan}
and explained using transition state theory in
Refs.~\cite{Godfrey2,Barnett}.  At high packing fractions
\begin{equation}
  \tau\sim\tau_0 \exp(\beta f \Delta_b)\,,
\label{taub}
\end{equation}
where $\tau_0$ is of the order of a disk collision time.  The argument
of the exponential in Eq.~\eqref{taub} can be understood from
Fig.~\ref{transitionstate}: $\beta f \Delta_b$ is the work done
against the piston in creating the extra length $\Delta_b$ in the
system which allows the defect to move.  In Ref.~\cite{Godfrey2} it
was shown that this extra length was $\Delta_b=\sqrt{4
  \sigma^2-h^2}-\sigma-\sqrt{\sigma-h^2}$, which can also be
understood by inspection of Fig.~\ref{transitionstate}.  The plateau
visible at larger packing fractions in Fig.~\ref{dy_squ} will end on
the time scale $\tau$.

\begin{figure}[ht]
  \centering
  \includegraphics[width=\columnwidth]{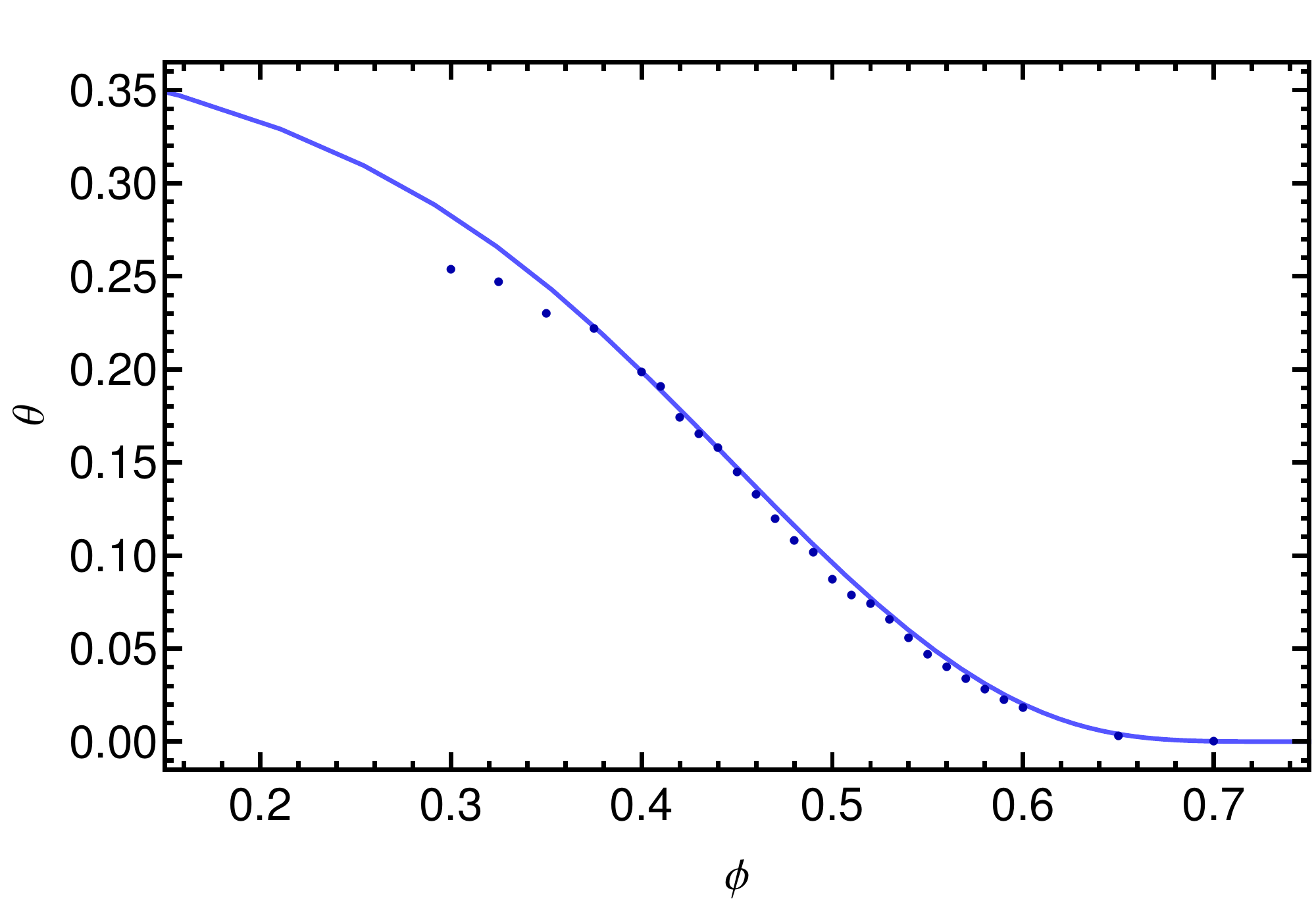} 
  \caption{(Color online) Variation of the defect density $\theta$
    with packing fraction~$\phi$.  The points give the observed defect
    densities in our simulations and the dashed line is the
    approximate theoretical relation predicted in
    Ref.~\cite{Godfrey2}.  The approximate relation is expected to
    improve in the limit of $\phi \rightarrow \phi_{\text{max}}$,
    which is seen here in the increased agreement between simulation
    and analytic values as $\phi$ increases, with the exception of the
    result for $\phi=0.7$ where we found 4 times more defects than
    expected.  This discrepancy is most likely due to poor
    equilibration as the trajectory was simulated for a time period
    much shorter than its relaxation time.  Relaxation times $\tau_D$
    are shown in Fig.~\ref{tau-D}.}
\label{theta}
\end{figure}

We now turn to the details of the diffusive behavior, indicated by the
dashed lines in Fig.~\ref{dy_squ}.  $\Delta^2(t)$ increases as
$\langle y_i(t)\, y_i(0) \rangle$ goes to zero.  The crossing of disks
from one side to the other of the channel is what drives this
correlation function to zero.  Fig.~\ref{transitionstate} shows that
this happens where there are defects in the zigzag arrangement of the
disks.  Let $\theta$ be the concentration of defects, so that the
number of defects is $N \theta$.  This number is readily determined in
numerical work using the procedure given in Ref.~\cite{Mahdi2}.
Figure~\ref{theta} shows $\theta$ as a function of packing fraction
$\phi$ obtained from the simulations and compared with the analytical
approach of Ref.~\cite{Godfrey2}, which becomes exact at large packing
fractions.
As time increases the number of disks flipped by the diffusion of the
defects will be of order $N \theta \sqrt{D t}$, where $D$ is the
diffusion coefficent of a defect.  It was argued in
Ref.~\cite{Godfrey2} that at large $f$, $D\tau_D \sim 1/\theta^2$.
Then
\begin{equation}
 \tau_D \sim \tau_0\exp(\beta f \Delta_c)\,,
\label{taualphastrong}
\end{equation}
where $\Delta_c=\sqrt{4 \sigma^2-h^2}+\sigma -3 \sqrt{\sigma^2-h^2}$.
The physical significance of the time scale $\tau_D$ is that it is the
time scale on which diffusing defects meet and
annihilate~\cite{Godfrey2}.  It appears to be the longest time scale
in the system -- the time scale for full equilibration.  Note that in the
diffusive region
\begin{equation}
 \Delta^2(t) \sim h^2 \sqrt{t/\tau_D}\,.
\label{sqrtt}
\end{equation}

\begin{figure}[ht]
  \centering
  \includegraphics[width=\columnwidth]{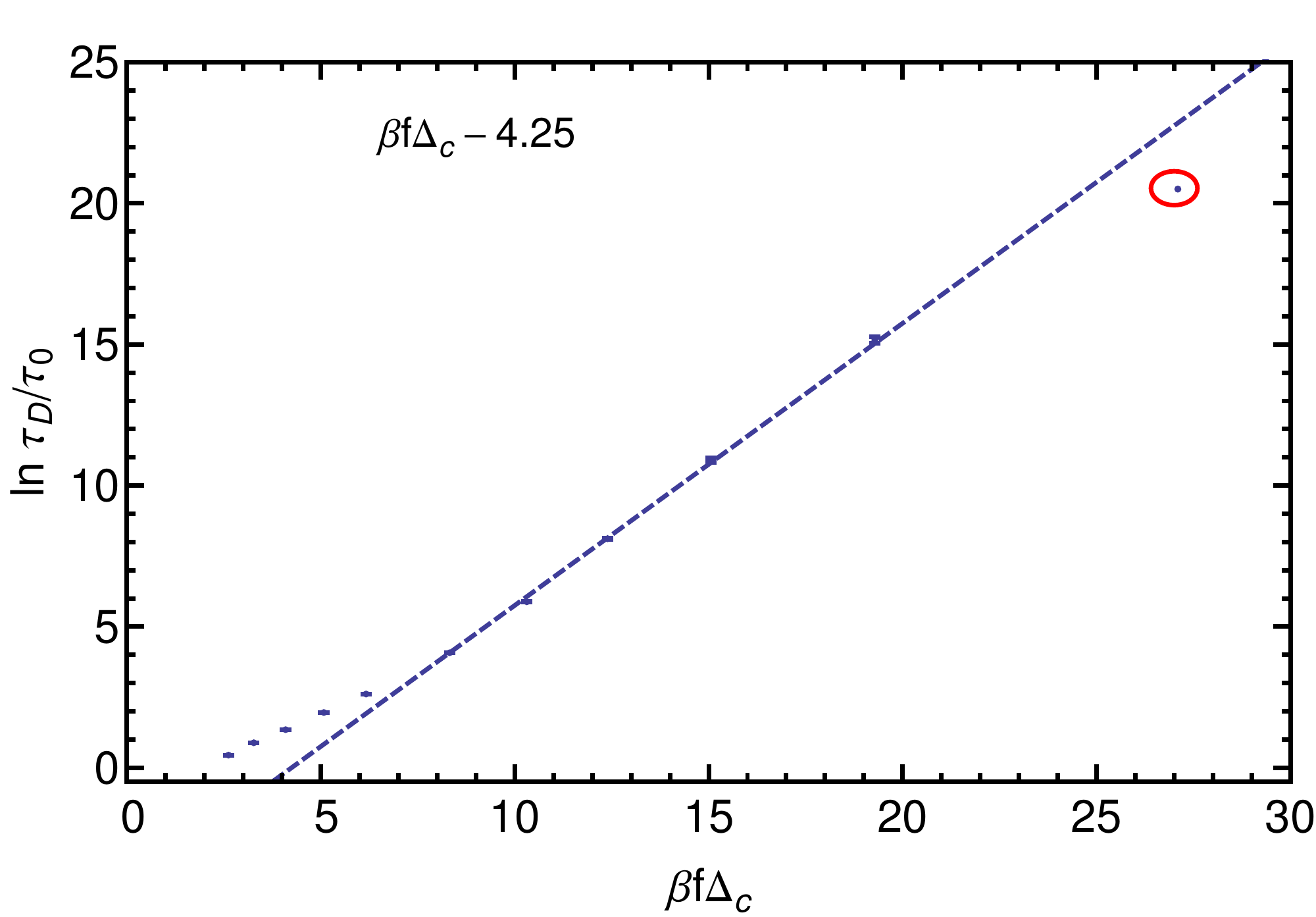} 
  \caption{(Color online) Rapid increase in the $\tau_D$-relaxation
    times with longitudinal force $f$, obtained by assigning a linear
    fit to the diffusive region of the mean square displacements (see
    Fig.~\ref{dy_squ}) and extrapolating to find the time where it
    meets the analytic final values of $\Delta^2(t)$.  The dashed line
    fits the exponential trend predicted by Eq.~\eqref{taualphastrong}
    that is $\ln \tau_D/\tau_0 \sim \beta f \Delta_c$ and becomes an
    increasingly better fit as $f$ becomes large, i.e. in the limit
    $\phi \to \phi_{\text{max}}$.  The circled data point,
    corresponding to $\phi=0.7$, deviates from this trend: as noted in
    the caption to~Fig.~\ref{theta}, the system failed to reach
    equilibrium for this value of~$\phi$.}
  \label{tau-D}
\end{figure}

By extending the dashed lines in Fig.~\ref{dy_squ} to the points where
they meet the analytic final values of $\Delta^2(t)$ we can estimate
values for~$\tau_D$.  Values of $\tau_D/\tau_0$ are plotted in
Fig.~\ref{tau-D}.  The collision timescale $\tau_0$ was estimated from
the mean collision rate per disk determined from our simulations, i.e.
\begin{equation}
  \frac1{\tau_0} = \frac{n_c}{N\tau_{\rm sim}}\,,
\end{equation}
where $n_c$ is the total number of collisions that occurred during the
simulation time~$\tau_{\rm sim}$.  The agreement with the prediction
\eqref{taualphastrong} is satisfactory for the results shown
in~Fig.~\ref{tau-D}.

\begin{figure}[ht]
  \centering
  \includegraphics[width=\columnwidth]{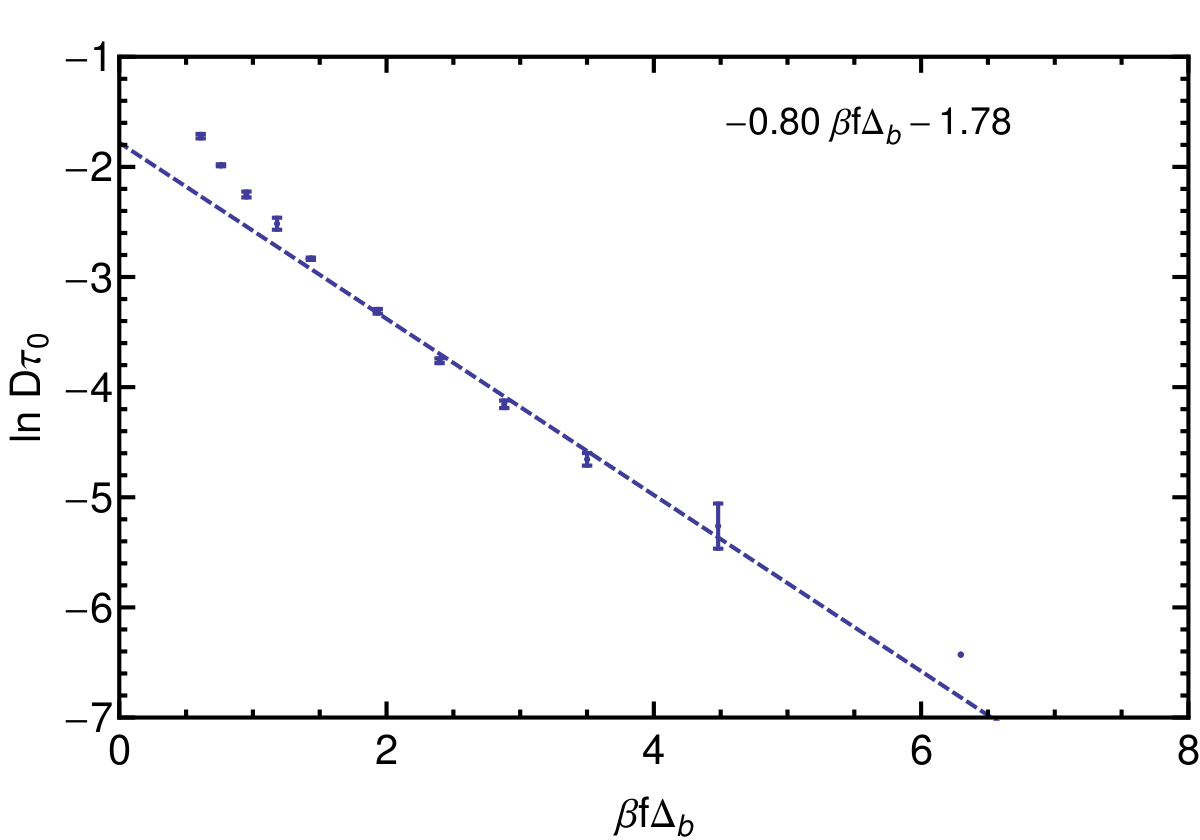} 
  \caption{(Color online) Variation of diffusion constant $D \propto
    1/\tau$ with longitudinal force $f$ using the $y$-intercept of
    linear fits to the diffusive regions shown in
    Fig.~\ref{dy_squ}. The dashed line fits the exponential trend
    predicted by Eq.~\eqref{taub}, $\ln \tau/\tau_0 \sim \beta f
    \Delta_b$.}
  \label{diff-coeff}
\end{figure}

We have also used the results shown in Fig.~\ref{dy_squ} to obtain the
diffusion coefficient for defects.  At high density, the diffusion
coefficient should be related to $\tau$ by $D\tau_0\sim\tau_0/\tau$,
and so might be expected to provide confirmation of Eq.~\eqref{taub}.
To find $D$, we make use of \eqref{sqrtt} in the form
\begin{equation}
  \Delta^2(t) \sim h^2\theta\sqrt{Dt}\,.
\label{thetaDt}
\end{equation}
Extrapolation of the dashed lines in Fig.~\ref{dy_squ} back to $t=1$
gives, via \eqref{thetaDt}, an estimate of $\theta\sqrt{D}$.  This in
turn provides $D$ when we make use of the values of $\theta$ found
from our simulations.  Results for $D\tau_0$ obtained in this way are
plotted in Fig.~\ref{diff-coeff}.  The results show a significant
departure from Eq.~\eqref{taub}, but there are several reasons why we
might expect our procedure to give poor results for the densities of
interest here.  First we note that for the smaller values of the
packing fraction the lines of slope $0.5$ in Fig.~\ref{dy_squ} are not
very convincing fits to the data: the linear portions of the curves
are very short.  Secondly, the mean spacing of defects given by
$1/\theta$ is not large (see Fig.~\ref{theta}) for $\phi$ in the range
0.4 to~0.6: interactions between defects may well modify the diffusion
coefficient significantly, leading to a $\theta$-dependent factor in
the relation $D\sim 1/\tau$.  Finally, as shown in
Ref.~\cite{Godfrey2}, $\theta$ itself is not a simple exponential
function of $\beta f\sigma$ at these moderate values of $\phi$, but
instead can be calculated quite accurately (see Fig.~\ref{theta}) from
a law of mass action.  Thus, even if $\tau$ has the activated form
\eqref{taub}, it is not certain that this can be ascertained from our
calculation of $D$ at the moderate densities accessible via our
simulation.

Fortunately, it is not necessary to rely on an estimate of the
diffusion coefficient to verify the activated behavior of~$\tau$.
Bowles and Saika-Voivod \cite{Ivan} have made a direct determination
of the channel-crossing time from their molecular dynamics simulation.
As shown in Ref.~\cite{Godfrey2}, their results are in satisfactory
agreement with Eq.~\eqref{taub}.

Finally we comment very briefly on the time-dependent correlation
function $R(t)=\langle y_i(t)\, y_i(0) \rangle$, mentioned in
Sec.~\ref{Introduction}.  $R(t)$ decays to zero because of the
diffusion of the defects in the regular zigzag arrangement of disks in
the channel.  Such a diffusive mechanism has been much studied
\cite{Redner} and leads to a stretched exponential decay $R(t) \sim
\exp(-\sqrt{t/\tau_D})$.  We have not attempted a direct verification
of this behavior of $R(t)$, as it is expected only for very large
times, $t\gg\tau_D$.

\subsection{Overlap correlations}
\label{sec:dyn-het}

We turn now to a study of overlap correlations, which correlate the
configuration of the system at time $t$ with a configuration drawn
from the equilibrium ensemble at time $t=0$.  Studies of such
correlations reveal the existence of dynamical heterogeneities in
three-dimensional glass-forming liquids.  It is therefore of interest
to see whether the system of disks in a channel shows similar
behavior.  In particular we shall make use of the (self) overlap
function
\begin{equation}
  Q(t)= \frac{1}{N}\sum_{i} w[y_i(t), y_i(0)]\,,
\label{Qdef}
\end{equation}
where $w[y_i(t),y_i(0)]=\frac{1}{2}(\sign(y_i(t)) \sign(y_i(0))+1)$ is
unity if disk $i$ is on the same side of the channel at times 0 and
$t$, and is zero otherwise; in the terminology of
Ref.~\cite{Berthier}, this quantity is the \emph{mobility} of
disk~$i$.  Similar overlap functions have been studied in three
dimensions: in that case, the overlap $w[\vec{r}_i,\vec{r}_j]$ has
been taken to be $1$ if $|\vec{r}_i-\vec{r}_j|\le 0.3\,\sigma$ and
zero otherwise~\cite{Glotzer,Royall0,Royall2}.  A significant
difference between this latter definition and our own is that our
overlap function does not depend on the $x$-coordinates of disks.
This modification eliminates an effect, specific to our
one-dimensional problem, of large fluctuations (${\sim}N^{1/2}$) in
the $x$-coordinates of disks.  These fluctuations can cause the
overlap between two configurations to be small even in cases where the
configurations would be identical, when described in terms of defects
in the zigzag arrangement of disks.

\begin{figure}[ht]
  \centering
  \includegraphics[width=\columnwidth]{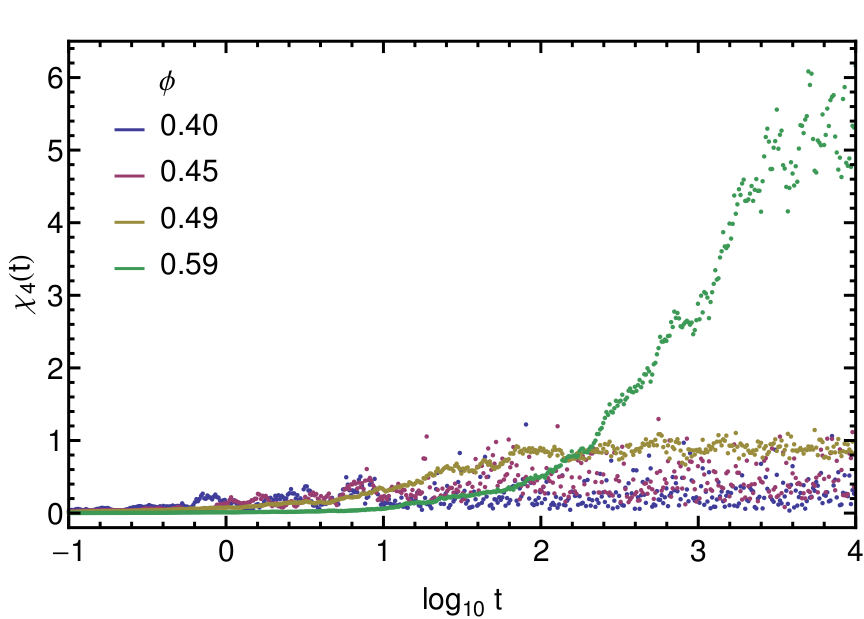} 
  \caption{(Color online) $\chi_4(t)$, as defined by
    Eq.~\eqref{chi4def}, as a function of time $t$ for various packing
    fractions $\phi$.  It approaches its largest values at a time
    corresponding to $\tau_D$. }
  \label{chi4}
\end{figure}

A quantity much studied for three-dimensional systems is the
four-point susceptibility $\chi_4(t)$, which is defined in terms of
the variance of $Q(t)$ via
\begin{equation}
  \chi_4(t)/N=\langle Q(t)^2 \rangle-\langle Q(t)\rangle^2.
\label{chi4def}
\end{equation}
This has been calculated for our one-dimensional system and is shown
in Fig.~\ref{chi4}.  In three dimensions it reaches a maximum on the
timescale $\tau_{\alpha}$ and subsequently decreases towards zero, but
for our system there is no decay back to zero.  This difference
between dimensions $d=3$ and $d=1$ has been discussed previously in
Ref.~\cite{Toninelli} and can be understood qualitatively as follows.
In three dimensions the self-overlap $w[\vec{r}_i(t),\vec{r}_i(0)]$
becomes small and is likely to remain small once a sphere has escaped
from its cage; it follows that the fluctuations in $Q(t)$ will also be
small for times that are long enough for most of the spheres to have
escaped from their cages.  This argument does not apply to our system
of disks in a channel because a disk can cross the channel any number
of times, so that $w[y_i(t),y_i(0)]$ is a fluctuating quantity of
order unity.  For any value of $t$, the $y$-coordinates of disks are
correlated over a range $\xi$ and are approximately independent over
larger distances.  From this we expect the contributions to $Q(t)$
from independent regions of size $\xi$ to be fluctuating quantities of
order $\xi/N$ for times $t>\tau_D$.  If the number of these regions is
$m_\xi\sim N/\xi$, we expect
\begin{equation}
  \chi_4/N=\operatorname{Var} Q(t) \sim m_\xi\left(\xi/N\right)^2 \,,
\label{xi4sim}
\end{equation}
which gives $\chi_4\sim\xi$ for long times.  This result can be
refined by using Eq.~\eqref{xidef} to evaluate the right-hand side of
\eqref{chi4def}; we obtain
\begin{equation}
  \chi_4(t) \approx \xi/4
\label{chi4longtime}
\end{equation}
for $t\gg\tau_D$ and $\xi\gg1$.  The last estimate
\eqref{chi4longtime} is consistent with the results shown in
Fig.~\ref{chi4} for $\phi=0.49$ and~$0.59$, for which $\xi=3.9$
and~$20.9$, respectively.

We have also determined the dynamical length scale from the four point
correlation function $\widetilde{S}_4(\kappa_x,t)$ defined as
\begin{equation}
  \widetilde{S}_4(\kappa_x,t) =
  \frac{1}{N}\langle\widetilde{Q}(\kappa_x,t)\,\widetilde{Q}^*(\kappa_x,t)\rangle\,,
\label{S_4de}
\end{equation}
where
\begin{equation}
  \widetilde{Q}(\kappa_x,t)=\sum_j e^{ - i\kappa_x j} y_j(t)\,y_j(0)
\end{equation}
and $\kappa_x=2 \pi m/N$, where $m=1$, 2, \dots, $N-1$.  The
$\kappa_x$-dependence of $\widetilde{S}_4(\kappa_x,t)$ follows a roughly
Lorentzian form, as found in Refs.~\cite{Glotzer,Royall2}, i.e.,
\begin{equation*}
  \widetilde{S}_4(\kappa_x,t)\approx
  \frac{A(t)}{1+\kappa_x^2\, \xi_4^2(t)}
\end{equation*}
for $\kappa_x\ne0$; the dynamical length scale $\xi_4(t)$ is obtained from
a fit to the simulation data.

\begin{figure}[ht]
  \centering
  \includegraphics[width=\columnwidth]{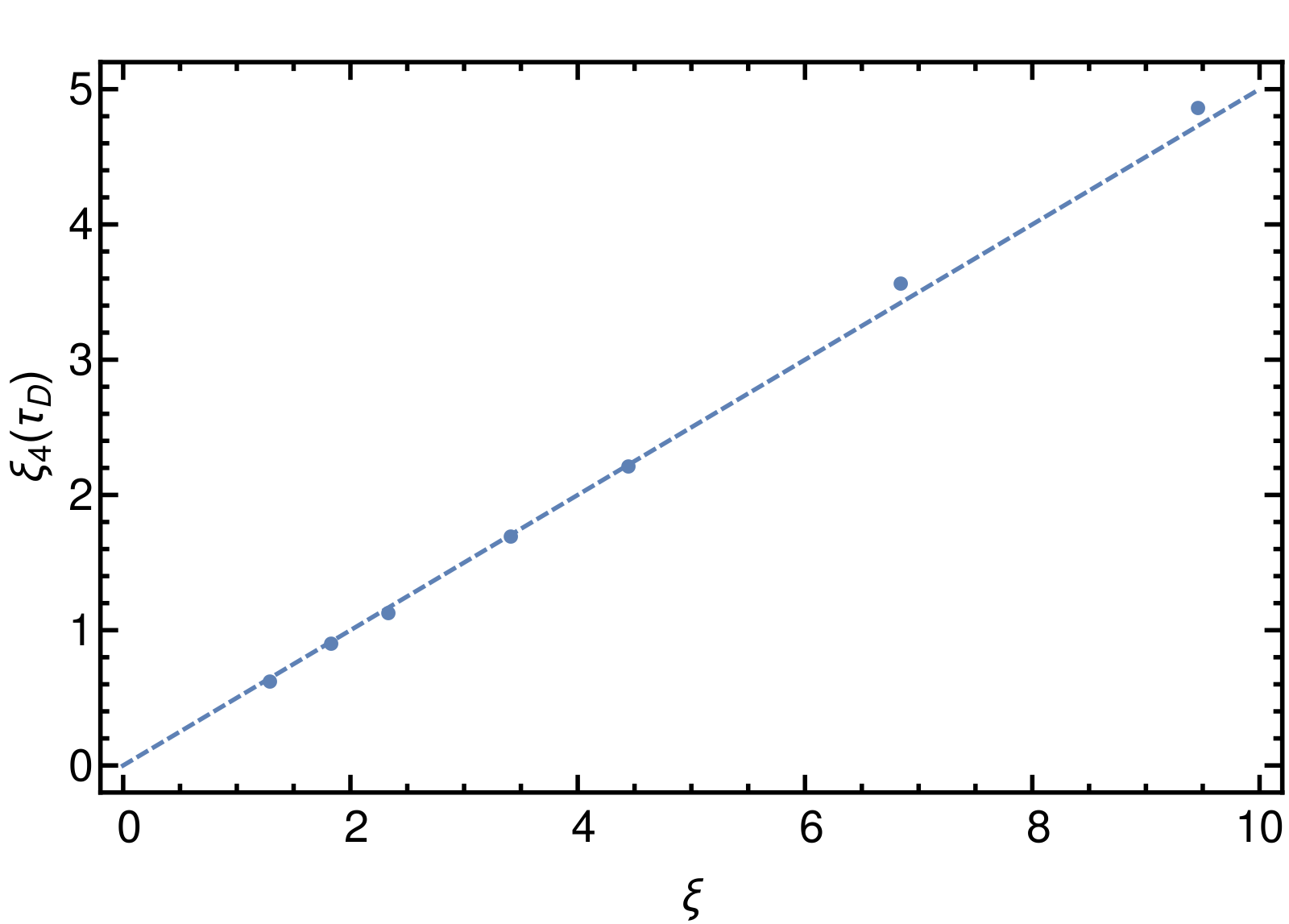} 
  \caption{(Color online) Plot of the dynamical length scale
    $\xi_4(\tau_D)$ versus the static length scale $\xi$ of zigzag
    order.  The gradient of the dashed line is $0.5$, corresponding to
    the predicted behavior, Eq.~\eqref{xi4-xi}.}
  \label{xi4tauD}
\end{figure}

We present results for $\xi_4(t)$ at two physically significant values
of~$t$.  At $t=\tau_D$ the system has reached equilibrium, so that
\begin{equation}
  \widetilde{S}_4(\kappa_x,\tau_D)\approx
  \frac{1}{N}\sum_{p,q} e^{-i\kappa_x (p-q)}\langle y_p\, y_q \rangle^2.
\label{S-4delimit}
\end{equation}
Then, by using Eq.~\eqref{xidef}, one finds the correlation length
\begin{equation}
  \xi_4(\tau_D)\approx\xi/2.
\label{xi4-xi}
\end{equation}
Here $\xi$ is the number of disks for
which zigzag order persists.  This dimensionless quantity is
calculated from the two largest eigenvalues, $\lambda_1$ and
$\lambda_2$, of the integral equation \eqref{integraleqKP}, using the
expression
\begin{equation}
  \xi = 1/{\ln(\lambda_1/|\lambda_2|)}\,,
\end{equation}
which follows from the spectral representation of correlation
functions, discussed for the case of the Ising model in Sec.~2.2 of
Ref.~\cite{Baxter} and applied to a system of hard disks in a
channel by Varga et al~\cite{VargaBalloGurin}.  The results
shown in Fig.~\ref{xi4tauD} are in good agreement with the
prediction~\eqref{xi4-xi}.

\begin{figure}[ht]
  \centering
  \includegraphics[width=\columnwidth]{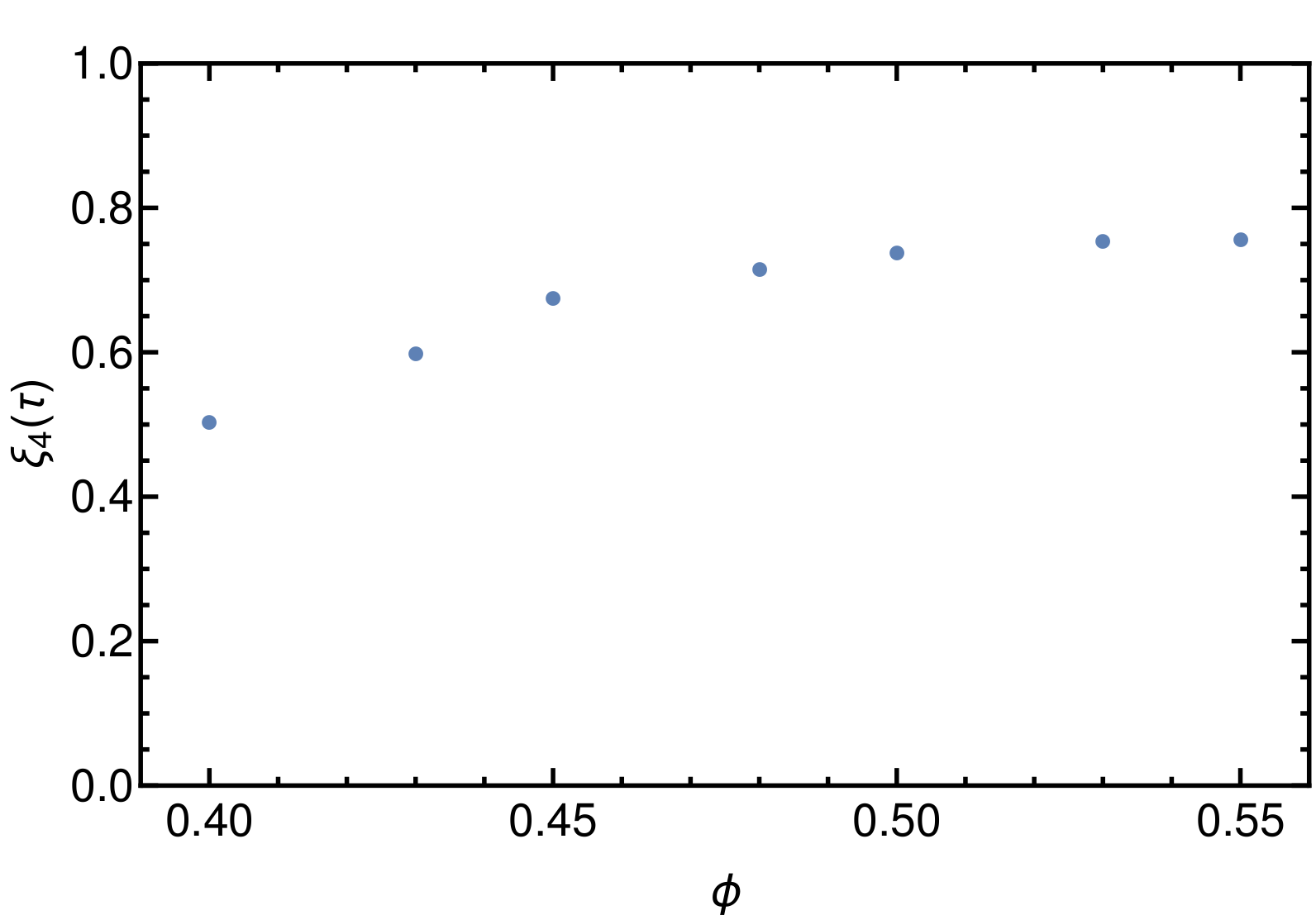} 
  \caption{(Color online) Plot of the dynamical length scale $\xi_4(
    \tau)$ versus the packing fraction $\phi$. }
  \label{xi4}
\end{figure}

If we choose for $t$ a value for less than the equilibration time
$\tau_D$, then $\xi_4(t)$ will be less than $\xi/2$.  On the time
scale $\tau$, the time for which particles are caged, $\xi_4(\tau)$ is
not proportional to $\xi$ and the results in Fig.~\ref{xi4} show it
instead to be of order $1$ and approximately independent of $\phi$ for
$\phi>\phi_d$.  This behavior of $\xi_4(\tau)$ is understandable, as
on the time scale $\tau$ the active regions are centered on the
defects, which involve O(1) disks.

The above studies show that our system has dynamical heterogeneities
of the kind expected in the defect-mediated scenario for glassy
dynamics~\cite{Toninelli,BB}.  In the packing-fraction regime where
the dynamics are activated, most of the disks will be largely frozen
except those in the vicinity of a defect.  The spacing between defects
is of order $\xi$ which can become very large.  The defects move about
and will eventually annihilate with each other.  This happens on the
time scale $\tau_D$, which is also the time scale on which new pairs
of defects are typically nucleated~\cite{Godfrey2}.

\section{Conclusions}
\label{conclusions}

There are some notable similarities between the behavior of our system
of disks in a narrow channel and that found for a fluid of hard
spheres in three dimensions.  In each of these systems at higher
densities the mean-squared displacement of particles has a plateau
that persists up to the characteristic time for the breaking of cages,
$\tau$ or $\tau_\alpha$.  Also, in both systems there is a crossover
from non-activated dynamics to activated dynamics as the density
increases and caging sets in.  In this latter respect, the behavior of
these systems is distinctly different from that of a hard-sphere
crystal with defects, where the particles are always caged and the
diffusion of defects will occur via an activated process.

There are, of course, some important differences.  In the channel
system, the growing bond-orientational order that we associate with
the caging of disks is clearly visible in the structure factor, and
this is not the case for the three-dimensional glass-forming liquids,
where higher-order correlation functions are required to reveal
bond-orientational order.  The difference here is due to the channel
walls, which form part of the cages: their direction is fixed and so
constrains the possible directions of the ``bonds'' between
nearest-neighbor disks.  This can be seen from the top and bottom
diagrams Fig.~\ref{transitionstate}, where only three nearest-neighbor
bond directions are possible at $\phi\to\phi_{\rm max}$.

Another difference appears in the behavior of the dynamical
susceptibility $\chi_4(t)$, which, for the channel system, does not
tend to zero for $t\to\infty$.  We note, however, that this is the
expected behavior of $\chi_4(t)$ in defect-mediated models of glassy
dynamics in one dimension~\cite{Toninelli}.

We believe that our work strongly supports the common idea
\cite{Royall,Royall0,Royall2,Tarjus,Liu} that glass behavior is a
consequence of geometry and the local arrangements around the
molecules in the supercooled liquid.  Our system is sufficiently
simple that we can quantitatively relate its dynamical features to
structural features.  The three-dimensional problem is much richer and
success along these lines is probably only just starting~\cite{Liu}.

The caging effects in our system mimic those seen in three dimensions.
These are normally modeled by mode-coupling theory~\cite{Gotze} and
are associated with a genuine dynamical transition, within that
approximation.  Our system is effectively one dimensional and is
unlikely to have a genuine phase transition.  We are skeptical that
the features that we see in the dynamics could be explained in any way
by mode-coupling arguments.  They seem instead to be more naturally
explained by dynamical processes associated with the developing
structural order in the system.  We suspect that the same might be
true of three-dimensional systems.

A noteworthy feature of the dynamics of our system of disks is the
existence of \emph{two} long time scales, $\tau$ and $\tau_D$.  In
three dimensions, only the analogue of $\tau$, which is the
cage-breaking time $\tau_\alpha$, is normally discussed.  But in two
dimensions a second, much longer, time scale has been revealed in
experimental studies of polydisperse colloidal crystals, reported by
Tanaka and co-workers~\cite{Tanakaetal,Tanaka}.  In their systems
there is a growing lengthscale $\xi$ for bond-orientational order, and
their second long timescale $\tau_\xi$ is associated with dynamical
correlations on that lengthscale.  This is strikingly similar to what
we have found for the case of disks in a channel, where the growing
lengthscale is that of zigzag ordering, which is a form of
bond-orientational order.  It might, of course, be that the second
long time scale is a feature of the dynamics only for one- and
two-dimensional systems.  One would expect two long timescales to
exist in three dimensions if escaping the cage could be associated
with moving a defect; the longer time scale would then be associated
with the time that the system needs to reach equilibrium via the
motion of defects.

\end{document}